\documentclass[journal=jacsat,manuscript=article]{achemso}
\usepackage{float}
\usepackage[version=3]{mhchem} 
\usepackage{hyperref}
\usepackage{cleveref}
\usepackage {graphicx}
\usepackage{braket}
\usepackage{titlesec}
\usepackage{xcolor}
\usepackage{ulem}

\usepackage{array}
\usepackage{makecell}
\usepackage{diagbox}
\usepackage{multirow}

\title{Unified Topological Dynamics of Merging Bound States in the Continuum for High-Order Topological Charges}

\author{Keren Wang}
\affiliation
{College of Physics, Sichuan University, Chengdu 610064, China}

\author{Lujun Huang}
\email{ljhuang@phy.ecnu.edu.cn}
\affiliation
{State Key Laboratory of Precision Spectroscopy, School of Physics, East China Normal University, Shanghai 200241, China}

\author{Wei Wang}
\email{w.wang@scu.edu.cn}
\affiliation
{College of Physics, Sichuan University, Chengdu 610064, China}

\begin{document}

\begin{abstract}
    Bound states in the continuum (BICs) are polarization singularities in momentum space whose topological charges (TCs) govern advanced light-matter interactions. While lattice symmetry protects the existence of robust BICs at the $\Gamma$-point (SP-BICs), it also restricts their TCs to low-order values. Achieving high-order TCs in common crystal lattices, such as $C_4$-symmetric systems, has therefore remained an open question. Here, we systematically demonstrate that high-order TCs that surpass fundamental symmetry bounds can be created through the rich dynamics of a parameter-driven merging process of off-$\Gamma$ BICs. We introduce a unified geometric framework based on the interplay between Fabry–Pérot interference and guided resonances, which uncovers different types of merging BICs dynamics, including near-isotropic, anisotropic, and cross-merging. Leveraging this mechanism, we realize unconventional TCs of up to $\pm3$ at either a symmetry-protected BIC or a degeneracy point in a simple $C_4$-symmetric photonic crystal slab. We further show that this high-order topology enables the generation of high-quality Bessel OAM beams, providing a physically transparent route toward engineering high-order topological photonics.
\end{abstract}
\maketitle

\section{Introduction}

Bound states in the continuum (BICs) have emerged as a unifying phenomenon across photonics, acoustics, and quantum systems, providing radiationless eigenmodes despite their coexistence with outgoing-wave channels~\cite{RN296,RN194,RN438,HUANG20231,Huang2024,OEA-2023-0186Nicolae}. In photonic crystal slabs (PCSs)~\cite{Joannopoulos2008PhotonicCM}, BICs naturally manifest as polarization singularities in momentum space, where the far-field polarization winds by an integer topological charge (TC) around each singular point~\cite{RN475}. Charge conservation under continuous deformations dictates the creation, motion, annihilation, and merging of BICs~\cite{RN248,RN246,RN240,RN697,RN695}. The frequency-resolved topological phase (TP) accumulated along an iso-frequency contour counts the enclosed TCs and directly controls geometric phases, spin–orbit interactions of light, beam shifts, lasing, and the generation of orbital angular momentum (OAM)~\cite{RN831,RN269,RN329,RN328}.

The symmetry-protected BIC (SP-BIC) is notably robust, yet its TC is bounded by lattice rotational symmetry: square ($C_4$) lattices allow only $q=\pm 1$, whereas hexagonal ($C_6$) lattices permit at most $q=-2$~\cite{RN831,RN246}. Achieving higher TC is highly desirable for obtaining larger TP and stronger geometric-phase modulation. Existing routes often rely on whispering-gallery–type resonators or quasicrystalline/moiré patterns~\cite{RN836,RN626,RN1156,RN322}; although effective, these approaches complicate design and fabrication and are less natural for compact planar PhC-slab platforms~\cite{RN461}. While low-symmetry photonic crystal slabs, in principle, hold promise for realizing higher-order topology with simpler architectures, systematic exploration in this direction remains surprisingly limited.

We note that a natural and conceptually attractive alternative is to leverage off-$\Gamma$ BICs—a broad family encompassing accidental (aBIC)\cite{RN1349,RN202}, Fabry–Pérot (FP)\cite{RN609,RN292,RN1041,advs.202200257}, and Friedrich–Wintgen (FW) BICs~\cite{RN722,RN818,RN238,Huang2024,AP.3.1.016004,Huang2021}. These labels emphasize specific interference mechanisms and partially overlap~\cite{RN150,RN438}, but from a geometric perspective they all represent polarization singularities that introduce additional TCs across the Brillouin zone~\cite{RN698,RN694}. By global charge conservation, off-$\Gamma$ singularities can move and merge, enabling redistribution of TC to or from the first Brillouin zone center. However, exploiting merging to realize higher TC requires a clear understanding of the underlying physical dynamics. Although merging-BIC phenomena were identified early on, and recent theoretical developments—including graph-theoretic descriptions~\cite{RN247} and perturbative analyses~\cite{RN736}—have provided valuable insight, these approaches mainly describe static charge configurations at fixed parameters. As a result, they offer only limited intuition into the dynamical origin of merging, the evolution paths of polarization singularities, and the conditions under which high-order vortices can form. Clarifying this dynamical picture remains a key challenge for advancing high-order topological photonics.

Here we present a unified and geometrically transparent mechanism based on FP interference of guided resonances (GRs). Longitudinal cavity modes couple to in-plane GRs and generate generalized off-$\Gamma$ BICs along high-symmetry directions, which appear as geometric intersections between FP-interference contours and the anisotropic GR-dispersion surface in momentum space. We further introduce anisotropic merging factors to quantify these merging geometries.  This framework shows how anisotropy enables iso-frequency loops to selectively enclose multiple same-sign off-$\Gamma$ charges, producing a large TP (for example $\pm3$) without creating a higher-order TC at $\Gamma$. It further uncovers a cross-merging process in which four off-$\Gamma$ charges converge at the $\Gamma$ point to form an emergent charge-3 ($|q|=3$) vortex. Using this mechanism, we realize TCs of up to $\pm 3$ at both the $\Gamma$-point SP-BIC and the degeneracy point (DP) in a $C_4$-symmetric PCS—exceeding the symmetry limits of robust SP-BICs—while maintaining a simple planar architecture. We show that the resulting high-order momentum-space topology directly maps into large geometric phase and clean high-order Bessel OAM beams~\cite{RN831,RN269,RN832}, providing a practical and fabrication-friendly route toward high-order topological photonic devices.

\section{Principle}

\begin{figure}[H]
    \centering
    \includegraphics[width=1\linewidth]{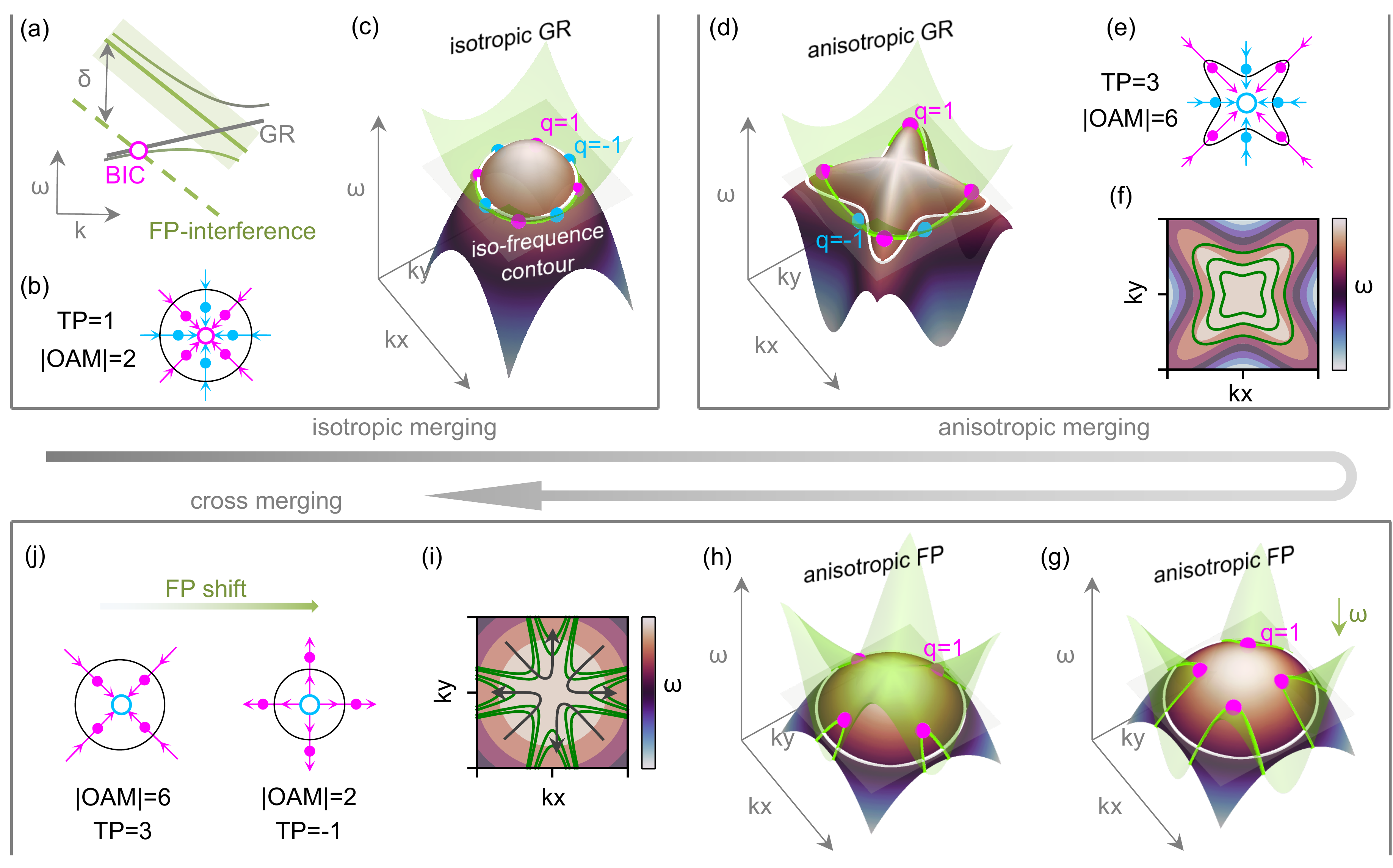}
    \caption{
\textbf{Geometric principle.}
(a) A minimal model in which FP interference competes with a GR through a single radiation channel, producing BICs. The BIC loci are captured by the geometric intersections between an effective FP-interference contour and the GR band.
(b) A representative, nearly uniform off-$\Gamma$ BIC distribution in momentum space and its evolution under a rigid shift of the FP-interference contour, illustrating the (isotropic) merging dynamics.
(c) The corresponding geometric model: off-$\Gamma$ BICs are mapped to the intersections between the FP-interference surface and the GR surface.
(d) Intersections formed when the GR dispersion is anisotropic while the FP-interference contour remains nearly isotropic.
(e) The associated BIC merging dynamics and
(f) the iso-frequency geometry enabling a high-order TP.
(g, h) Intersections when the FP-interference contour possesses a saddle-like anisotropy against a nearly isotropic GR.
(i, k) The resulting cross-merging dynamics in which BIC trajectories switch between symmetry directions, yielding high-order TC.
}
    \label{fig:principle}
\end{figure}

As illustrated in \autoref{fig:principle}, the origin of off-$\Gamma$ BICs in photonic crystals can be understood from the pattern of FP interference. The interplay between an FP mode and a GR coupled to a single radiation channel is captured by the non-Hermitian effective Hamiltonian $H_{\rm eff}$~\cite{RN150, RN719}:
\begin{equation}
H_{\rm eff} =
\begin{pmatrix}
\omega_1 & \kappa \\
\kappa & \omega_2
\end{pmatrix}
-i
\begin{pmatrix}
\gamma_1 & \pm \sqrt{\gamma_1\gamma_2} \\
\pm \sqrt{\gamma_1\gamma_2} & \gamma_2
\end{pmatrix},
\label{eq:Heff}
\end{equation}
where $\omega_{1}-i\gamma_{1}$ and $\omega_{2}-i\gamma_{2}$ are the complex eigenfrequencies of the isolated GR and FP modes, respectively. With a single open channel, a BIC occurs when the radiative amplitudes destructively interfere, which imposes
\begin{equation}
\kappa(\gamma_1-\gamma_2)=\pm\sqrt{\gamma_1\gamma_2}\,(\omega_1-\omega_2).
\label{eq:BIC_condi}
\end{equation}
Hence, for a given detuning $\delta=\omega_1-\omega_2$, the BIC forms on the GR branch. Equivalently, one may shift the FP dispersion by $\delta$ and regard the intersections between the shifted FP curve and the GR as directly specifying the BIC positions on the GR. We refer to the shifted FP curve as the FP-interference contour (light green in the schematics), see \autoref{fig:principle}(a).

To remain general, we focus on square lattices with $C_4$ symmetry, which guarantees
$\omega(\mathbf{k}){=}\omega(-\mathbf{k})$ and
$\omega(k_x,k_y){=}\omega(k_y,k_x){=}\omega(-k_x,k_y){=}\omega(k_x,-k_y)$.~\cite{RN246}
In the sketches we use a simple parabolic model obeying these symmetries (explicit expressions are given in the Supplementary Material Sec.~1~\cite{supp1}). In typical situations the FP-interference contour intersects the GR surface along an almost circular curve (green). Because linear-polarization channels decouple on the Brillouin-zone (BZ) high-symmetry lines, the BIC condition is met and eight off-$\Gamma$ BICs emerge, symmetrically arranged along the principal directions~\cite{RN240}. By charge conservation and symmetry, their TC alternate as $+1$ and $-1$; we denote them by purple and cyan, respectively~\cite{RN247}.

The TC is defined as the winding number of the far-field polarization angle $\phi(\mathbf{k})$ around a polarization singularity. We use\cite{RN140, RN247}
\begin{equation}
q=\frac{1}{2\pi}\oint \nabla_{\mathbf{k}}\phi(\mathbf{k})\cdot d\mathbf{k},
\label{eq:TC}
\end{equation}
and analogously define a frequency-resolved TP on an iso-frequency contour $C(\omega)$ of the GR band,
\begin{equation}
Q_{\rm TP}=\frac{1}{2\pi}\oint_{C(\omega)}\nabla_{\mathbf{k}}\phi(\mathbf{k})\cdot d\mathbf{k}
=\sum_{i\in\mathrm{inside}(C)} q_i .
\label{eq:TP}
\end{equation}
Thus $Q_{\rm TP}$ is obtained by counting the total TC enclosed by $C(\omega)$. As a concrete implication, under circular excitation the cross-polarized channel accumulates a Pancharatnam--Berry phase $\exp[-i\,2Q_{\rm TP}\,\varphi]$, generating vortex beams with $\mathrm{OAM}=\pm 2Q_{\rm TP}$.~\cite{RN269} Although we refer to a ``phase'', the TP is the cumulative winding of the polarization vortex. In the symmetric scenario of \autoref{fig:principle}(a), any choice of $\omega$ encloses TCs that cancel pairwise, preventing the build-up of large $Q_{\rm TP}$.

In realistic structures the FP-interference contour can itself be dispersive, and the GR dispersion along the $\Gamma$--X and $\Gamma$--M directions can differ. Their intersections therefore exhibit rich geometry, opening routes to generate higher TP and TC. We highlight two canonical cases:

\paragraph*{Case I: anisotropic GR, nearly isotropic FP-interference.}
When only the GR dispersion is anisotropic [\autoref{fig:principle}(d)], the iso-frequency contour of the GR and the FP-interference contour can form a closed, anisotropic intersection. The eight off-$\Gamma$ BICs are then strongly staggered along two BZ directions. By selecting $\omega$ within an appropriate window, the iso-frequency loop may enclose four off-$\Gamma$ BICs of the same charge, yielding an additional $\pm 4$ contribution to $Q_{\rm TP}$. Full-wave simulations further show that this contribution is partially canceled by the charge at $\Gamma$, resulting in a net $Q_{\rm TP}=\pm 3$. In this regime all eight BICs still merge at $\Gamma$ as the FP-interference contour is shifted (anisotropic merging akin to \autoref{fig:principle}(b)), so no higher-order TC is created; this regime is therefore characterized as a high-order TP without high-order TC [\autoref{fig:principle}(e, f)].

\paragraph*{Case II: saddle-like FP-interference, nearly isotropic GR.}
When the FP-interference contour carries a saddle anisotropy while the GR is nearly isotropic [\autoref{fig:principle}(g, h)], the intersections become open. In this geometry four off-$\Gamma$ BICs of identical charge remain within the BZ while the opposite-charge partners can be regarded as having moved to infinity. Upon shifting the FP-interference contour, the BIC trajectories switch between the $\Gamma$--X and $\Gamma$--M directions and undergo a cross-merging at $\Gamma$ that yields a charge-$3$ BIC, as depicted in \autoref{fig:principle}(i, k)\cite{RN467, Li2025DegenerateBIC}. Iso-frequency selection is then more flexible: any loop enclosing these four same-sign off-$\Gamma$ BICs produces a high-order TC and, concomitantly, an enhanced TP. The high-order phases---together with the anisotropic merging and cross-merging processes and the associated path switching---will be validated in simulations.

\section{Numerical Results}

\begin{figure}[h]
    \centering
    \includegraphics[width=1\linewidth]{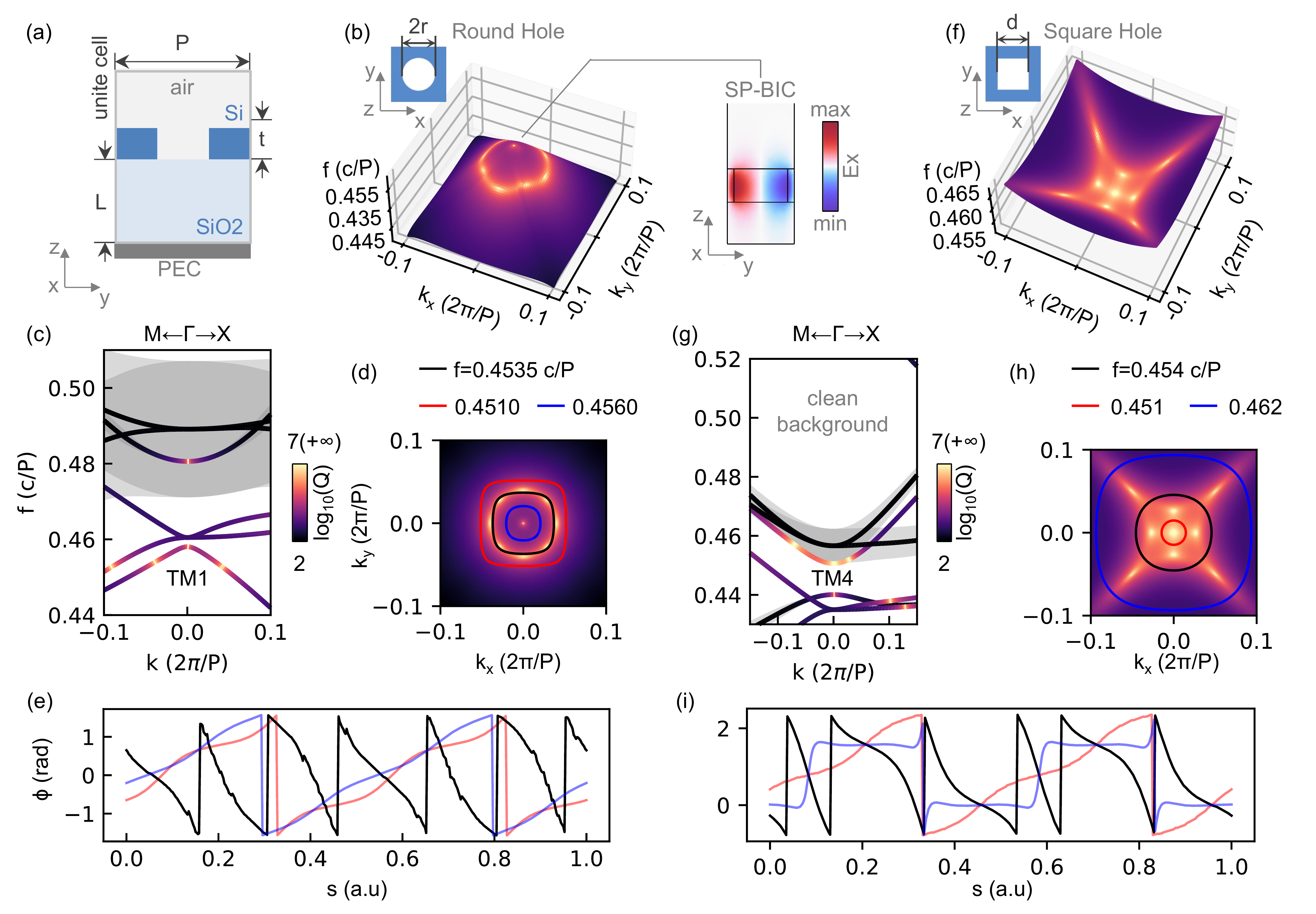}
    \caption{
    \textbf{Anisotropic merging and high-order TP without high-order TC.}
    (a) Schematic of the PCS–FP cavity. The left and right panels correspond to circular-hole (structure~A) and square-hole (structure~B) configurations, respectively.
    (b) For structure A, the TM\textsubscript{1} band in 2D momentum space colored by the radiative quality factor $Q$. Right inset: near field of the central symmetry-protected BIC (SP-BIC).
    (c) 1D dispersion of structure A along M–$\Gamma$–X.
    (d) Three representative iso-frequency intersections (solid lines) overlaid on the $Q$-factor heat map.
    (e) Polarization angle $\phi$ versus normalized arclength $s$ traced along the three contours in (d).
    (f–i) Same as (b–e) but for structure B (TM\textsubscript{4} band).
    }
    \label{fig:aniso_merging}
\end{figure}

To conveniently tune the position of the FP-interference contour, we employ a composite platform consisting of a PCS above an FP cavity; see \autoref{fig:aniso_merging}(a). The structure comprises a PCS of thickness $t$ and refractive index $n=3.5$, separated by a subwavelength spacer of thickness $L$ and index $n=1.45$ from a metallic mirror modeled by a perfect electric conductor (PEC) boundary. The PCS and the mirror naturally form an FP cavity; varying $L$ shifts the FP mode frequency and, hence, the FP-interference contour. The in-plane geometry is defined by the lattice constant $P$ and either a circular hole of radius $r$ (structure~A) or a square hole of side length $d$ (structure~B). Such structures are readily fabricated via CVD and standard electron-beam lithography~\cite{wang2025}. We emphasize that our theory also applies to all-dielectric platforms without a mirror, where the FP mode can be captured by an effective-index model; these systems exhibit similar merging dynamics~\cite{RN694, Li2025DegenerateBIC, RN467, RN240}, while our mirror-based realization offers a more transparent geometric interpretation.

\paragraph*{High-order TP from anisotropic merging.}
We first demonstrate the generation of high-order TP, wherein anisotropic merging leads to iso-frequency contours that selectively enclose only one type of off-$\Gamma$ BICs, thereby accumulating a large TP without creating a higher-order TC at $\Gamma$.
In $C_4$ photonic crystals, the lowest-order TM band (TM\textsubscript{1}) exhibits the well-known square anisotropy\cite{Joannopoulos2008PhotonicCM}. We therefore begin with TM\textsubscript{1} on structure A with parameters $(P,t,r,L)=(400~\mathrm{nm},\,200~\mathrm{nm},\,0.4P,\,250~\mathrm{nm})$.
The 2D dispersion in \autoref{fig:aniso_merging}(b) shows an annular region of large $Q$ around the central SP-BIC—this ring traces the intersections between the FP-interference contour and the GR surface. Off-$\Gamma$ BICs emerge on the high-symmetry lines ($\Gamma$–X and $\Gamma$–M), which we denote as X-BIC and M-BIC, respectively.

The M–$\Gamma$–X cut in \autoref{fig:aniso_merging}(c) reveals that TM\textsubscript{1} disperses more weakly along $\Gamma$–M than along $\Gamma$–X. The FP mode lies around $f=0.48$–$0.50$ $c/P$, so the FP–GR intersections shift the M-BICs to slightly higher frequencies than the X-BICs. Consequently, for frequencies between these two sets, the iso-frequency contour encloses only the four M-BICs while excluding the X-BICs [\autoref{fig:aniso_merging}(d)].
With $q(\text{SP-BIC})=+1$ and $q(\text{M-BIC})=-1$ (opposite to X-BIC), the contour at $f=0.4535\,c/P$ (black curve) accumulates
$Q_{\rm TP}=-1\times4+1=-3$, as shown in \autoref{fig:aniso_merging}(e).
By contrast, contours at $f=0.4510\,c/P$ (red) and $f=0.4560\,c/P$ (blue) enclose opposite-charge sets and yield $Q_{\rm TP}=+1$.

Replacing the circular holes with square holes enhances the anisotropy of the FP interference and therefore amplifies the anisotropic-merging effect.\cite{adfm.202309982} In \autoref{fig:aniso_merging}(f–i), structure B with $(P,t,d,L)=(400~\mathrm{nm},\,200~\mathrm{nm},\,0.6P,\,180~\mathrm{nm})$ exhibits a similar annular high-$Q$ intersection on the TM\textsubscript{4} band, together with unequal M- and X-BIC frequencies [\autoref{fig:aniso_merging}(g)]. Although TM\textsubscript{4} itself is not strongly anisotropic, the FP interference (around $f=0.46$–$0.48~c/P$) is, which drives a pronounced anisotropic merging on TM\textsubscript{4}. A sizable frequency window (e.g., $f=0.454\,c/P$, black curves in \autoref{fig:aniso_merging}(h, i)) then yields $Q_{\rm TP}=-3$. Because the GR anisotropy is weaker here, the iso-frequency contour is more circular, which is advantageous for applications such as vortex-beam generation (see \autoref{fig:cross_conversion}). Notably, in \autoref{fig:aniso_merging}(g) there are no other optical modes in the range $f=0.48$–$0.52~c/P$, allowing the FP–GR interaction to be well approximated as an effective two-level system. The central role of FP anisotropy is further verified by polarization-resolved analysis in Supplementary Sec.~4~\cite{supp1}. We stress that the high-order TP in this regime results from selectively enclosing a subset of TCs; hence it exists only within specific frequency intervals, otherwise opposite charges contaminate the integral.
\color{black}
\paragraph*{Anisotropic merging factors.}
The above examples indicate that, although the off-$\Gamma$ BICs in structure~A are nearly isotropic in momentum space, they exhibit a slight anisotropy in frequency, whereas in structure~B the off-$\Gamma$ BICs are anisotropic in both momentum and frequency.
To quantify how differently the off-$\Gamma$ BICs approach $\Gamma$ along distinct high-symmetry directions, we introduce anisotropic merging factors defined in both frequency and in-plane momentum space.

Let $f_{\Gamma}$ be the band frequency at $\Gamma$, and let $f_{1}$ ($f_{2}$) denote the frequency of the off-$\Gamma$ BIC located on the high-symmetry ray $\Gamma\!\rightarrow\!1$ ($\Gamma\!\rightarrow\!2$).
We define the frequency detunings $\Delta f_{i}$ and the frequency merging factor $F_f$:
\begin{equation}
\Delta f_{i} \equiv f_{i}-f_{\Gamma},
\qquad
F_f \equiv \frac{\Delta f_{1}}{\Delta f_{2}} .
\label{eq:Ff_ratio}
\end{equation}
For $C_4$ lattices we choose $(1,2)=(X,M)$, i.e. $F_f=\Delta f_{X}/\Delta f_{M}$, while for $C_6$ lattices we can choose $(1,2)=(K,M)$.
Similarly, let $\mathbf{k}_{i}$ be the in-plane wavevector of the off-$\Gamma$ BIC on $\Gamma\!\rightarrow\!i$ and define
\begin{equation}
\Delta k_{i} \equiv \left|\mathbf{k}_{i}-\mathbf{k}_{\Gamma}\right|=\left|\mathbf{k}_{i}\right| ,
\qquad
F_k \equiv \frac{\Delta k_{1}}{\Delta k_{2}} .
\label{eq:Fk_ratio}
\end{equation}
A graphical illustration of $\Delta f_i$ and $\Delta k_i$ is provided in Supplementary Material Sec.~2\cite{supp1}.
We note that the frequency anisotropy can be directly accessed under monochromatic excitation, thereby providing an experimentally convenient handle for achieving large TP.

Simultaneous merging of the off-$\Gamma$ BICs along the two symmetry directions implies $F_f\!\to\!1$ and $F_k\!\to\!1$.
Notably, in cross-merging, the absence of an off-$\Gamma$ BIC along one symmetry direction renders the anisotropic merging factor singular, so that it evolves between $0$ and $+\infty$ as the BIC is transferred between high-symmetry directions.
The detailed interpretation of the sign and singular behavior of the merging factors is discussed in Supplementary Material Sec.~2\cite{supp1}.
Under these definitions, structure~A exhibits weak anisotropic merging ($F_f=1.18$, $F_k=1.00$), whereas structure~B shows much stronger anisotropy ($F_f=0.18$, $F_k=0.36$). Detailed tracking of the off-$\Gamma$ BIC trajectories as functions of $L$ confirms that, for both structures~A and~B, $F_f$ and $F_k$ ultimately converge to unity at the merging point (see Supplementary Sec.~3~\cite{supp1}).
\color{black}

\begin{figure}[h]
    \centering
    \includegraphics[width=1\linewidth]{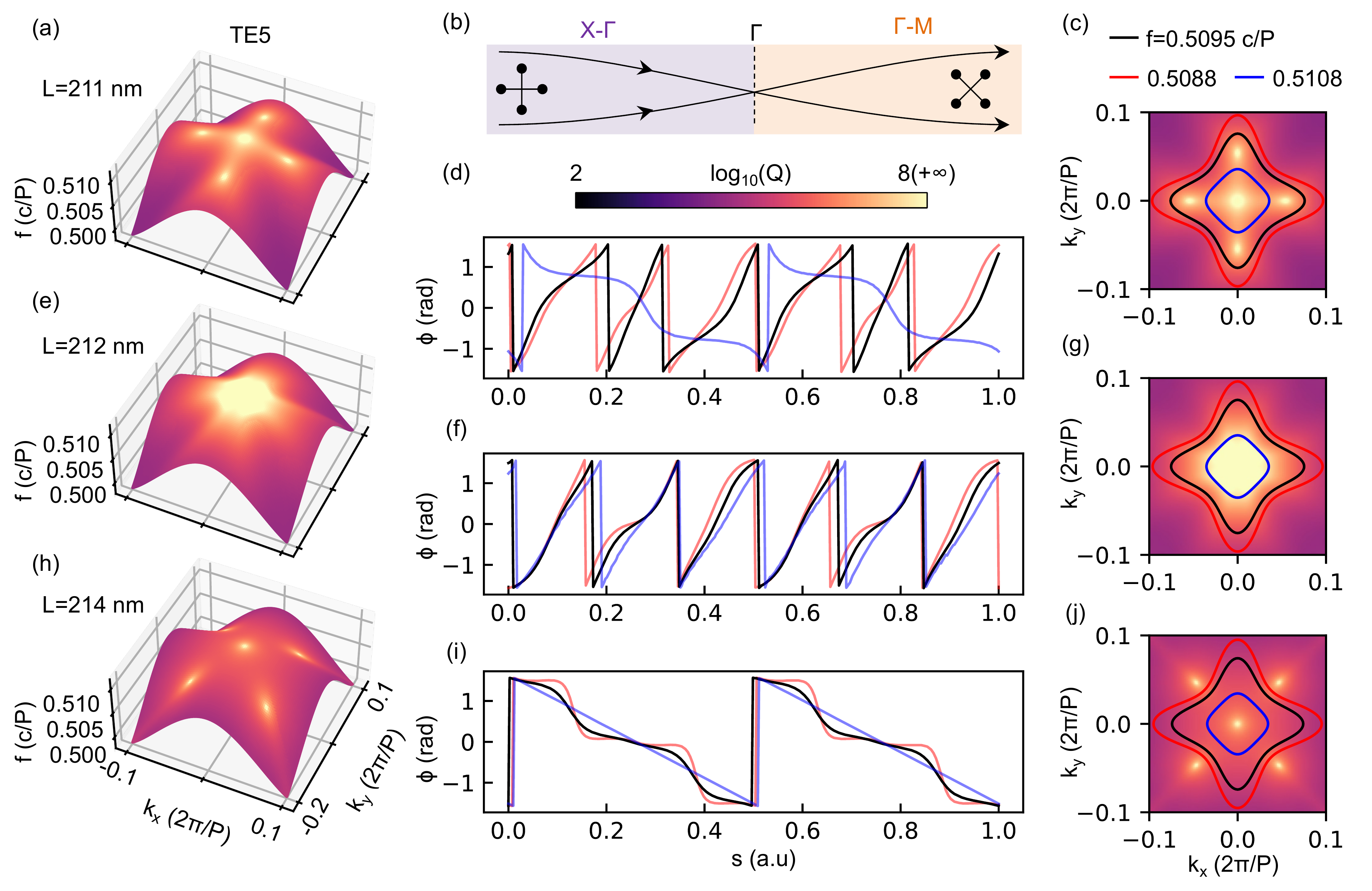}
    \caption{
    \textbf{Cross-merging at a BIC: creation of high-order TC.}
    (a, e, h) 2D $Q$-colored dispersion of TE\textsubscript{5} for structures C\textsubscript{1}, C\textsubscript{2}, C\textsubscript{3} with different FP thicknesses $L$.
    (b) Schematic of cross-merging: four M-BICs convert into four X-BICs through a charge-$3$ super-BIC at $\Gamma$.
    (c, g, j) Iso-frequency contours (solid) overlaid on the $Q$-map for C\textsubscript{1}, C\textsubscript{2}, C\textsubscript{3}.
    (d, f, i) Polarization angle $\phi$ versus normalized arclength $s$ along the three contours in (c,g,j).
    }
    \label{fig:cross_merging-BIC}
\end{figure}

\paragraph*{High-order TC from cross-merging.}

We now demonstrate strict high-order TC formation—without any contamination from opposite charges—via cross-merging. Such high-order TC supports broad-band high-order TP. We show that the merging dynamics follow our geometric theory and can occur either at a BIC or at a DP.

For the BIC case (\autoref{fig:cross_merging-BIC}), structure C uses $(P,t,r)=(400~\mathrm{nm},\,200~\mathrm{nm},\,0.3P)$ and varies $L$ across C\textsubscript{1}, C\textsubscript{2}, C\textsubscript{3} with $L{=}\,211,\,212,\,214~\mathrm{nm}$. In C\textsubscript{1}, only four M-BICs are present near $\Gamma$ [\autoref{fig:cross_merging-BIC}(a)]. Increasing $L$ moves them toward the central SP-BIC and, at $L\!\approx\!212~\mathrm{nm}$ [\autoref{fig:cross_merging-BIC}(e)], they cross-merge at $\Gamma$ into a charge-$3$ super-BIC. Further increasing $L$ [\autoref{fig:cross_merging-BIC}(h)] splits the super-BIC into four X-BICs. This evolution, summarized in \autoref{fig:cross_merging-BIC}(b), matches the scenario of \autoref{fig:principle}(i, j) and yields a high-order TC that, in turn, enables broadband high-order TP. 
Indeed, after the formation of the super-BIC (at $L{=}212~\mathrm{nm}$), all three iso-frequency contours in \autoref{fig:cross_merging-BIC}(f, g) produce $Q_{\rm TP}=+3$, whereas for $L{=}211$ or $214~\mathrm{nm}$ only sufficiently low frequencies that enclose four off-$\Gamma$ BICs achieve high-order TP [\autoref{fig:cross_merging-BIC}(d, c, i, j)]. Compared with the anisotropic-merging regime of \autoref{fig:aniso_merging}, cross-merging thus relaxes the frequency selectivity.
We also note that, to the best of our knowledge, such a large positive TC has not been reported in quasicrystalline structures~\cite{RN322}.
\textcolor{black}{From the perspective of the anisotropic merging factor defined above, cross-merging here is characterized by a singular behavior, evolving between $0$ and $+\infty$, in stark contrast to anisotropic merging.}

\begin{figure}[h]
    \centering
    \includegraphics[width=1\linewidth]{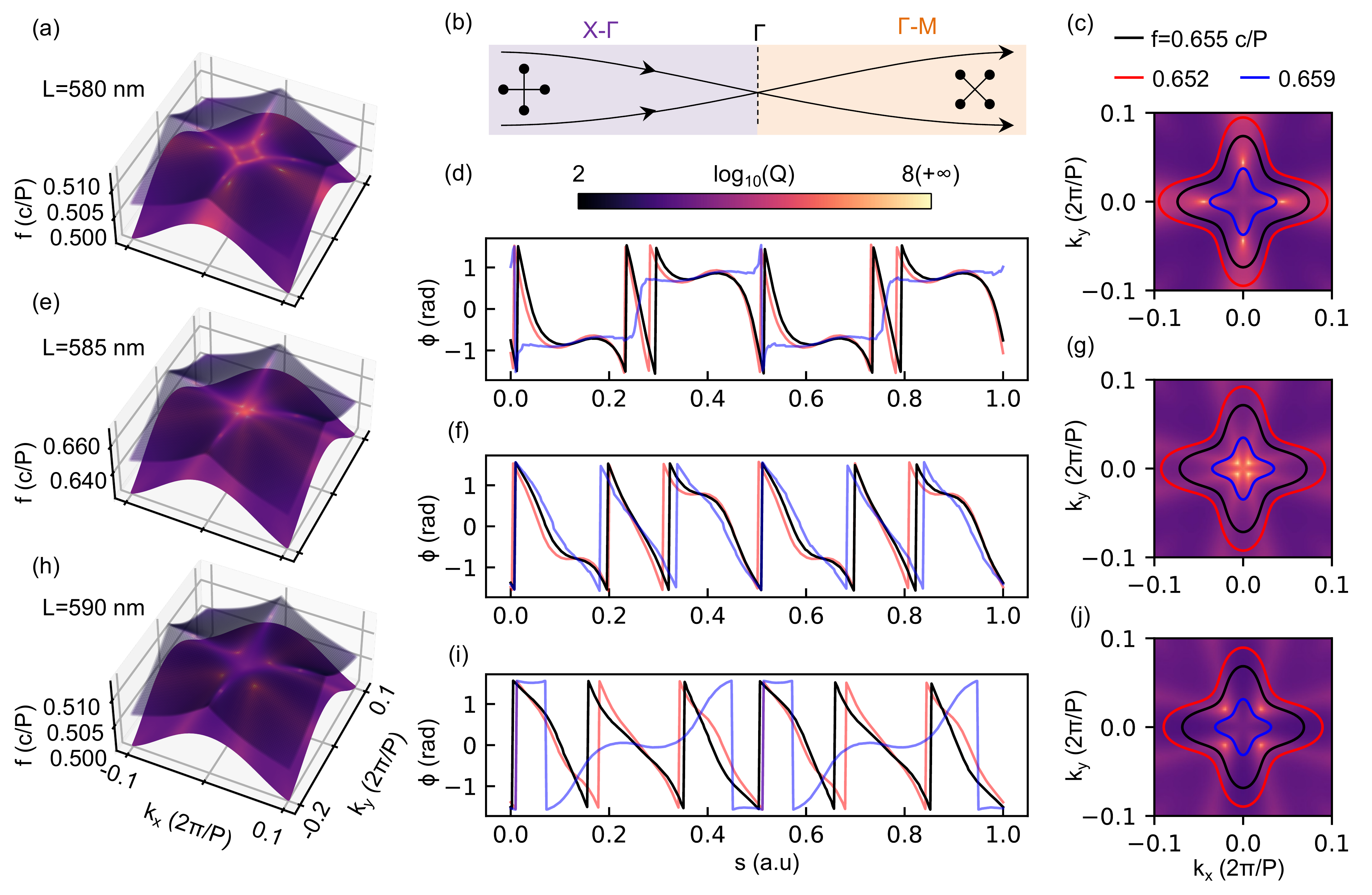}
    \caption{
        \textbf{Cross-merging at a DP: high-order TC without $Q$-divergence.}
        (a,e,h) 2D $Q$-colored dispersion of TE\textsubscript{5} for structures D\textsubscript{1}, D\textsubscript{2}, D\textsubscript{3} with different FP thicknesses $L$.
        (b) Schematic of cross-merging across a DP.
        (c, g, j) Iso-frequency contours (solid) overlaid on the $Q$-map for D\textsubscript{1}, D\textsubscript{2}, D\textsubscript{3}.
        (d, f, i) Polarization angle $\phi$ versus normalized arclength $s$ along the three contours in (c, g, j).
    }
    \label{fig:cross_merging-DP}
\end{figure}

For the DP case (\autoref{fig:cross_merging-DP}), structure D uses $(P,t,r)=(1300~\mathrm{nm},\,200~\mathrm{nm},\,0.3P)$ and $L{=}\,580,\,585,\,590~\mathrm{nm}$ for D\textsubscript{1}, D\textsubscript{2}, D\textsubscript{3}. We focus on the lower band near the DP. As $L$ varies, the lower band follows the same cross-merging sequence as in the BIC case [compare \autoref{fig:cross_merging-DP}(a, e, h) with the schematic in \autoref{fig:cross_merging-DP}(b)]. Because the DP is the locus of a trivial band touching in a $C_4$ system~\cite{RN317,RN694}, both lower and upper bands exhibit analogous cross-merging and are capable of forming high-order TC; one may predict that eight same-sign off-$\Gamma$ BICs co-merge in the complementary band~\cite{Li2025DegenerateBIC}. (Details for the upper band are given in the Supplementary Material Sec.~5~\cite{supp1}.) In $C_4$ lattices, the DP itself carries a well-defined polarization topological charge while its radiative $Q$ remains finite; at the critical thickness where four same-sign off-$\Gamma$ BICs coalesce, the central feature becomes a charge-$3$ super-BIC with $Q\!\to\!\infty$, reverting to a finite-$Q$ DP away from this point. Consistent with this picture, \autoref{fig:cross_merging-DP}(c, g, j) and \autoref{fig:cross_merging-DP}(d, f, i) show that structure D supports $-3$ TC and the associated $Q_{\rm TP}=-3$, demonstrating that the merging mechanism is not limited to $+3$.

\begin{figure}[h]
    \centering
    \includegraphics[width=1\linewidth]{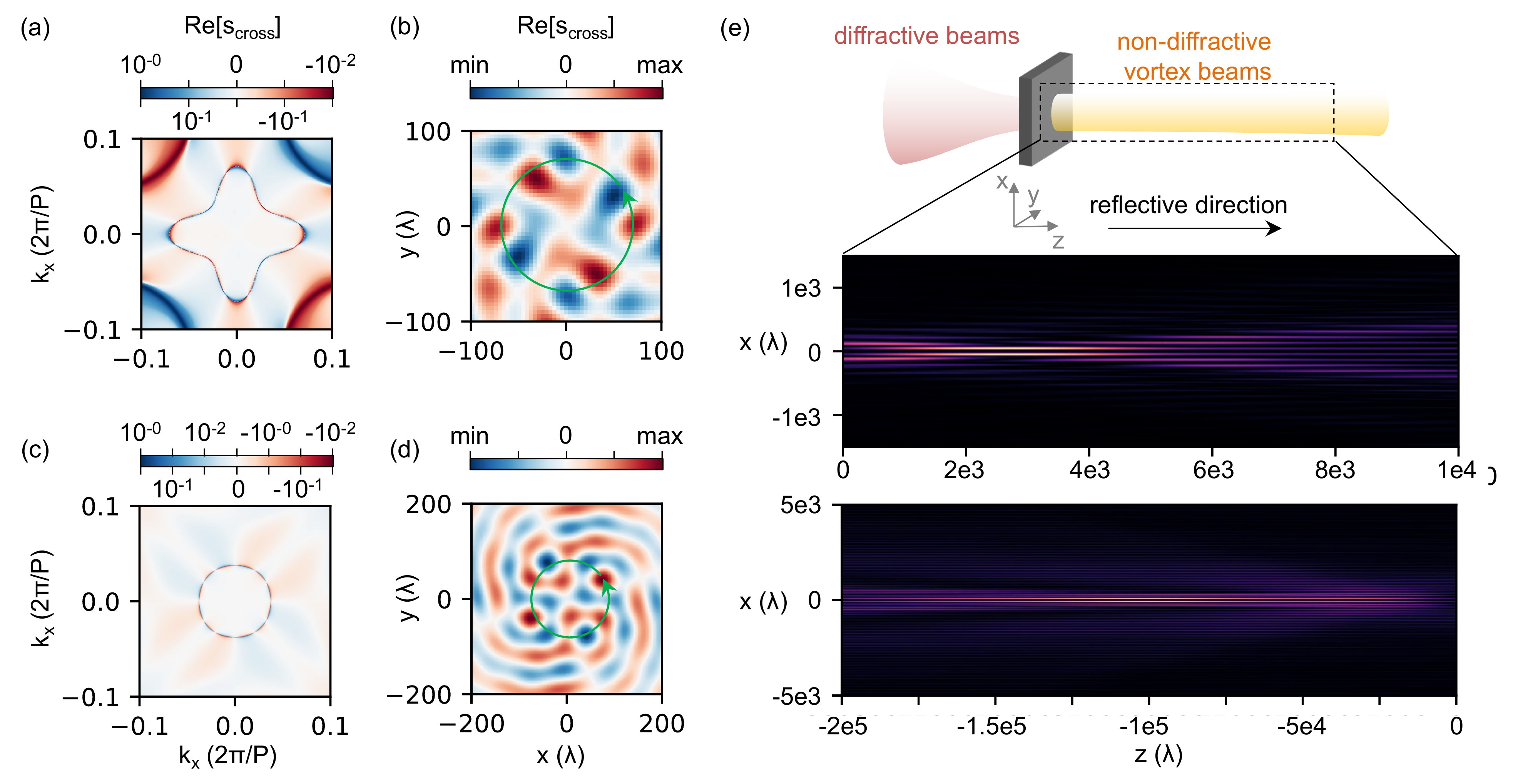}
    \caption{
    \textbf{High-order OAM and non-diffracting beams.}
    (a,b) For structure D\textsubscript{2} under circular excitation at $f=0.655\,c/P$, the complex cross-polarized reflection amplitude $s_{\rm cross}$ in momentum space (a, signed-log scale) and its real-space field obtained by Fourier transform (b).
    (c,d) Same as (a,b) but for structure B at $f=0.453\,c/P$.
    (e) Schematic of non-diffracting beams produced by a $C_4$ periodic structure; the two axial intensity profiles correspond to (a,b) and (c,d), respectively.
    }
    \label{fig:cross_conversion}
\end{figure}

\paragraph*{Vortex beams with high-order OAM and non-diffracting propagation.}
Finally, we show that the high-order TP enables high-order OAM in the reflected cross-polarized channel. In a PCS, the OAM is set by the TP accumulated along the iso-frequency contour, and the OAM sign flips with the handedness of the incident circular polarization; we therefore focus on $|{\rm OAM}|$.
We compare structures B and D. For structure D at $f=0.655\,c/P$, the momentum-space pattern of $s_{\rm cross}$ in \autoref{fig:cross_conversion}(a) exhibits twelve sign changes in $\mathrm{Re}\,s_{\rm cross}$, indicating $|{\rm OAM}|=6$, which is confirmed by the real-space Fourier transform in \autoref{fig:cross_conversion}(b) showing a high-order phase vortex. Owing to the intrinsic GR anisotropy, the momentum-space profile is strongly petal, which is suboptimal for beam quality. By contrast, structure B features a more circular iso-frequency contour; at $f=0.453\,c/P$, the analytically extracted momentum-space vortex in \autoref{fig:cross_conversion}(c) is more isotropic. Its real-space Fourier transform, obtained after discarding high-$|k|$ components ($|k|>0.02\times2\pi/P$), corresponding to an aperture-limited flat-top beam, yields a cleaner high-order vortex [\autoref{fig:cross_conversion}(d)]. This comparison highlights the practicality of our approach for generating large-$|{\rm OAM}|$ beams with high spatial quality. See Supplementary Material Sec.~7 for the analytical method used~\cite{supp1}. From the perspective of skyrmionic textures, higher-order TC also corresponds to higher-order polarization textures.~\cite{RN1322, RN1149}

Near a merging BIC, the large-$Q$ resonance narrows the momentum-space spectrum, which translates into extended non-diffracting propagation in real space~\cite{RN544}. Moreover, the beam acquires directionality set by the sign of the modal dispersion~\cite{Kim2025};  In our two examples, the group velocities are negative (structure~D) and positive (structure~B), respectively, leading to co- and counter-directed propagation, as confirmed by Fourier-optics simulations [\autoref{fig:cross_conversion}(e)].

\section{Discussion and Conclusion}

The FP–GR intersection picture offers a compact geometric interpretation of how off-$\Gamma$ BICs are generated and evolve. It unifies the descriptions of FW-, FP-, and accidental BICs under a single interference framework~\cite{RN150}: the FP contour determines where destructive interference occurs, while the GR dispersion controls how these singularities move and merge. Within this view, anisotropic merging explains the formation of high-order TP without increasing the central TC, whereas a saddle-like FP contour drives cross-merging that produces a charge-$3$ super-BIC and broad-band high-order TP. 
\textcolor{black}{Beyond the specific systems studied here, the anisotropic merging factors introduced in this work provide a simple and transferable tool for characterizing BIC merging geometries in a wide range of photonic structures, which can be readily extracted from band-structure calculations or experiments.
Importantly, anisotropic merging and cross-merging are not isolated scenarios but can be continuously connected through parameter tuning. As demonstrated by structure~F in Supplementary Sec.~3~\cite{supp1}, a configuration that appears to exhibit anisotropic-merging behavior is, upon closer inspection, revealed to be a cross-merging process, in which BIC trajectories pass through $\Gamma$ and ultimately merge at finite off-$\Gamma$ momenta.
We note that the simple geometric model presented in Fig.~1, which approximates the relevant dispersions as quadratic surfaces, does not directly apply to structure~F. Nevertheless, the observed evolution remains fully consistent with the generalized FP–GR interference picture, in which the merging behavior is governed by the overall geometry of the FP–GR intersections in momentum space, rather than by the specific details of a minimal parabolic model.}

An interesting question concerns the maximum topological charge that can exist at the $\Gamma$ point in PCS with finite symmetry. We summarize in Supplementary Material Sec.~6 the momentum-space far-field topological charges reported to date in finite-symmetry PCSs~\cite{supp1}. On the one hand, symmetry constrains in momentum space indicate that, for example, a $C_4$-symmetric system permits only odd TCs at $\Gamma$~\cite{RN475}. On the other hand, the analysis of symmetry-protected eigenmodes shows that the stable TCs enforced by $C_4$ symmetry are restricted to $0$ and $\pm 1$~\cite{RN831}. Together, these results imply that the charge-$3$ super-BIC formed through merging is not symmetry protected; instead, it behaves more like an accidental high-TC singularity in parameter space. As discussed in Supplementary Sec.~5~\cite{supp1}, the near-field patterns of the $\Gamma$-point mode (whether at the DP or at the BIC) exhibit no significant change throughout the cross-merging process. This is fully consistent with the expectation that high-order TCs arising from merging lack parameter robustness and therefore do not impose symmetry-protected field profiles.

At this point, it remains possible in principle to obtain even larger TCs through more elaborate merging processes or other parameter-driven mechanisms. However, such TCs would likewise not be symmetry protected and would generally require fine tuning of structural parameters. Whether finite-symmetry PCS can host robust, high-order $\Gamma$-point vortices beyond the known symmetry limits thus remains an open question and an intriguing direction for future exploration.

In summary, we establish a unified and physically transparent mechanism for understanding and engineering merging-BIC dynamics. The FP-interference model reveals how anisotropy and cavity detuning can control the creation of high-order TC and TP, enabling compact, planar generation of high-quality vortex beams and offering a general strategy for manipulating topological polarization states in photonic systems~\cite{RN501, RN682}.

\section*{Acknowledgments}
This work was supported by the Natural Science Foundation of Sichuan Province (Grant No. 2024NSFSC0460) and the National Natural Science Foundation of China (Grant No.12474377).


\section*{Competing interests}
The authors declare no competing financial interests.

\section*{Supplementary information}
See Supplementary Material for supporting content.

\newpage

\setcounter{section}{0}
\setcounter{subsection}{0}
\setcounter{subsubsection}{0}
\setcounter{figure}{0}
\setcounter{table}{0}
\setcounter{equation}{0}


\renewcommand{\thesection}{S\arabic{section}}

\begin{center}
{\Huge \bfseries Supplementary Material\par}

\vspace{36pt}

\end{center}

\vspace{12pt}

\begingroup
\small
\setlength{\parskip}{0pt}
\setlength{\baselineskip}{10pt}
\tableofcontents
\endgroup

\newpage

\section{Simple Parabolic Band for Geometry Modeling}
We consider a square-lattice photonic crystal with point-group symmetry $C_4$.
For a singly-degenerate band with dispersion $\omega(\mathbf{k})$ in the in-plane Bloch momentum
$\mathbf{k} = (k_x,k_y)$, the $C_4$ symmetry implies the following relations:
\begin{equation}
\omega(k_x,k_y)
=
\omega(-k_x,k_y)
=
\omega(k_x,-k_y)
=
\omega(k_y,k_x)
=
\omega(-k_y,k_x).
\label{eq:C4_symm}
\end{equation}
These constraints encode inversion symmetry $\mathbf{k}\to-\mathbf{k}$, mirror symmetries with respect
to the $\Gamma$--X and $\Gamma$--Y axes, and fourfold rotational symmetry.
It is therefore convenient to parametrize $\mathbf{k}$ by polar coordinates
\begin{equation}
r^2 = k_x^2 + k_y^2, 
\qquad
\theta = \arctan2(k_y,k_x),
\label{eq:polar}
\end{equation}
and to expand the dispersion in invariants of $C_4$.
To lowest nontrivial order, the angular dependence enters through $\cos(4\theta)$,
which is the leading Fourier component compatible with Eq.~(\ref{eq:C4_symm}).

In the geometric model used in the main text, both the guided-resonance (GR) band and the
FP-interference contour are approximated as parabolic dispersions along each high-symmetry direction.
This corresponds to a quadratic expansion in $r$ with the simplest $C_4$-symmetric angular dependence.
We therefore write a generic surface as
\begin{equation}
E(\mathbf{k})
=
\alpha\, r^2 + \beta\, r^2 \cos(4\theta),
\label{eq:generic_surface}
\end{equation}
where $\alpha$ controls the isotropic parabolic curvature and $\beta$ controls the fourfold anisotropy.
Because $\cos(4\theta)=1$ along the $\Gamma$--X and $\Gamma$--Y directions
($\theta = 0,\pi/2,\pi,3\pi/2$),
while $\cos(4\theta)=-1$ along the $\Gamma$--M directions
($\theta = \pi/4,3\pi/4,5\pi/4,7\pi/4$),
Eq.~(\ref{eq:generic_surface}) yields parabolic dispersions along each
high-symmetry line, with different effective curvatures.
This realizes precisely the situation described in the main text:
the band is locally parabolic in $k$ along each symmetry direction,
but the curvature can differ between $\Gamma$--X and $\Gamma$--M due to anisotropy.

Within this framework, the BIC condition derived from the effective Hamiltonian
[Eq.~(3) in the main text] can be implemented geometrically as the intersection between
the GR surface $E_2(\mathbf{k})$ and the FP-interference surface $E_3(\mathbf{k})$.
For a fixed choice of parameters, the BIC loci on the GR branch are determined by the solutions of
\begin{equation}
E_2(\mathbf{k}) = E_3(\mathbf{k}),
\label{eq:BIC_geom_condition}
\end{equation}
or explicitly,
\begin{equation}
r^2
\Bigl[
C_{2}^{\mathrm{iso}} - C_{3}^{\mathrm{iso}}
+ C_{2}^{\mathrm{aniso}} \cos(4\theta)
- C_{3}^{\mathrm{aniso}} \cos(4\theta + \varphi_3)
\Bigr]
= \Delta.
\label{eq:BIC_geom_explicit}
\end{equation}
The number and arrangement of solutions of Eq.~(\ref{eq:BIC_geom_explicit})
fix the positions of the off-$\Gamma$ BICs in momentum space.
For small anisotropy, the intersections form an almost circular contour,
yielding eight off-$\Gamma$ BICs distributed uniformly along the principal directions.
As the anisotropic coefficients and phase are varied,
the intersection geometry becomes strongly distorted or even open,
leading to the anisotropic merging and cross-merging trajectories described in the main text,
and enabling the emergence of higher-order topological charges and large topological phases.

This minimal, symmetry-consistent model suffices to reproduce all the qualitative
features of off-$\Gamma$ BIC formation, anisotropic merging, cross-merging, and
the associated high-order polarization topology reported in the main text.

\newpage

\section{Interpretation of anisotropic merging factors}.

\begin{figure}[H]
    \centering
    \includegraphics[width=0.3\linewidth]{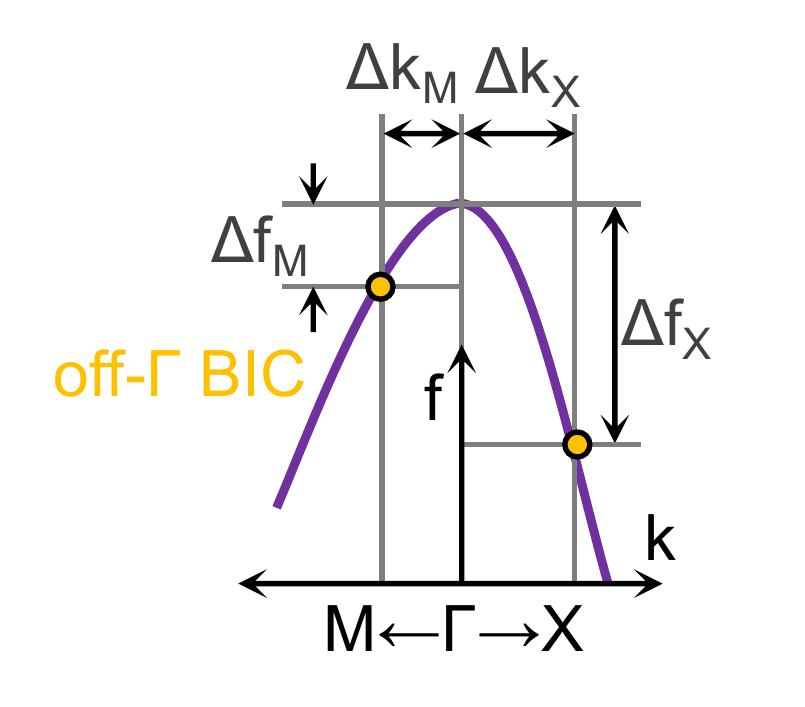}
    \caption{
    A graphical illustration of $\Delta f_i$ and $\Delta k_i$, defined in the main text.
    }
    \label{figS:square_hole}
\end{figure}

We provides brief remarks on the anisotropic merging factors introduced in the main text. First, the sign of the frequency factor $F_f$ reflects the local band geometry. A negative $F_f$ indicates opposite frequency slopes along the two symmetry directions, corresponding to a saddle-like dispersion near $\Gamma$. In this case the iso-frequency contour is open, and selective enclosure of off-$\Gamma$ BICs is no longer well defined. Second, in cross-merging scenarios off-$\Gamma$ BICs are absent along one symmetry direction over part of the tuning range. Accordingly, one detuning $\Delta f_i$ (or $\Delta k_i$) vanishes while the other remains finite, causing the anisotropic merging factors to become singular and evolve between $0$ and $\infty$. This behavior reflects symmetry-direction switching of BIC trajectories, rather than a divergence of any physical observable. Finally, we note that the anisotropic merging factors are not topological invariants. They depend on the chosen symmetry directions and serve only as diagnostic quantities for distinguishing anisotropic merging from cross-merging.

\section{An intermediate structure bridging anisotropic merging and cross-merging}

\begin{figure}[H]
    \centering
    \includegraphics[width=0.85\linewidth]{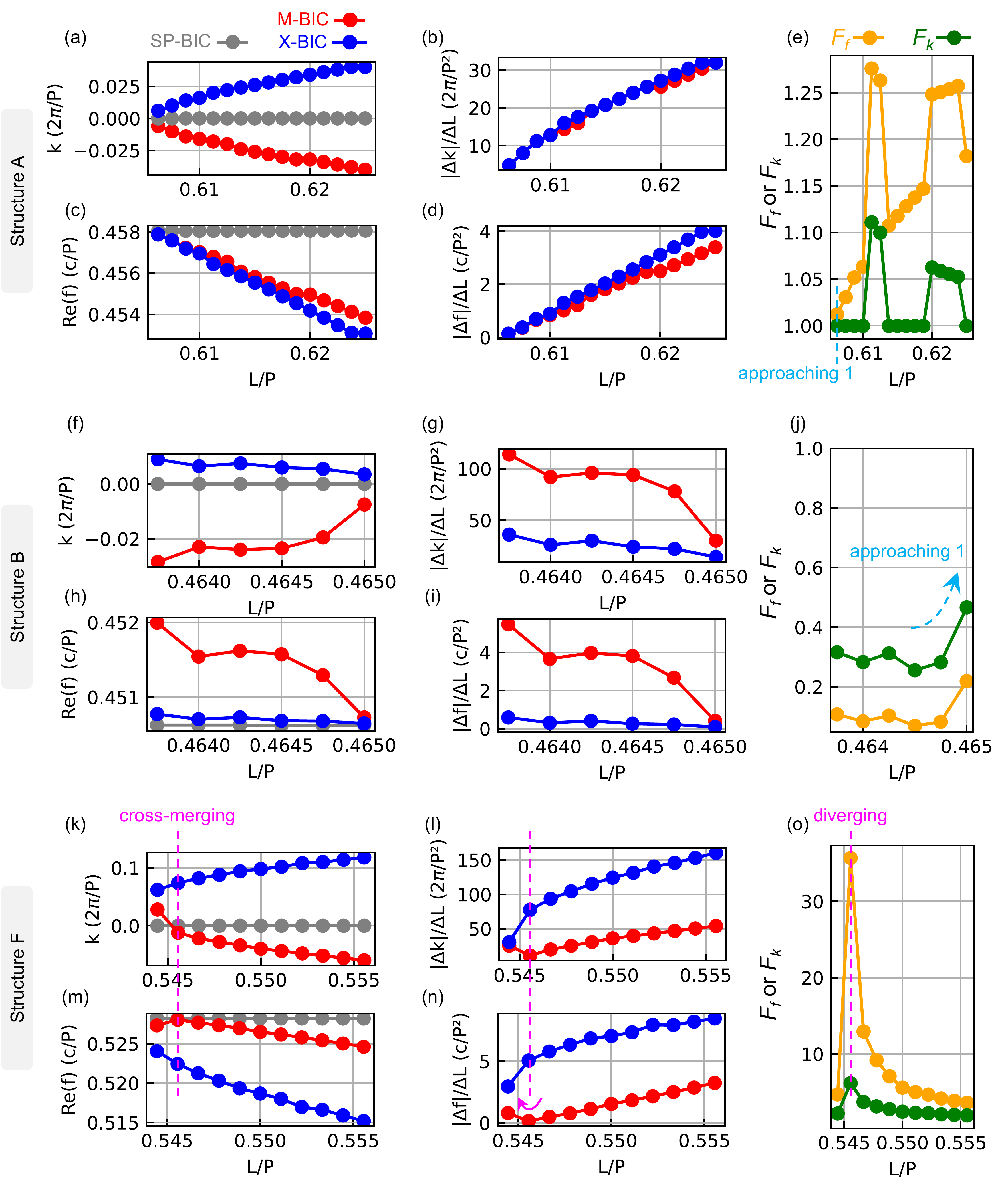}
    \caption{
    Tracking of off-$\Gamma$ BIC trajectories as the FP thickness $L$ is tuned for structures~A, B, and F. Panels show the evolution of BIC wave vectors, eigenfrequencies, velocities, and anisotropic merging factors. Structure~F highlights symmetry-direction transfer and off-$\Gamma$ merging, bridging anisotropic merging and cross-merging.
    }
    \label{figS:merging_tracking}
\end{figure}

In the main text, anisotropic merging and cross-merging are presented as two distinct regimes: in the former, off-$\Gamma$ charges merge simultaneously at $\Gamma$, whereas in the latter they switch symmetry directions and effectively pass through $\Gamma$. To clarify the connection between these processes, we explicitly track the evolution of off-$\Gamma$ BICs as a function of the FP-cavity thickness $L$, by uniformly sampling both the in-plane wave vector and $L$, as summarized in \autoref{figS:merging_tracking}.

For structure~A, we record the momentum positions and eigenfrequencies of the off-$\Gamma$ BICs [\autoref{figS:merging_tracking}(a,c)], together with their velocities in momentum and frequency spaces, defined as $|\Delta k|/\Delta L$ and $|\Delta f|/\Delta L$ [\autoref{figS:merging_tracking}(b,d)]. M-BICs are marked in red, X-BICs in blue, and the $\Gamma$-point SP-BIC in gray. Using the anisotropic merging factors defined in Supplementary Sec.~2, we further extract the evolution of $F_k$ and $F_f$ with $L$ [\autoref{figS:merging_tracking}(e)]. Minor discontinuities arise from the finite momentum resolution ($10^{-4}$ in units of $2\pi/P$). Since BICs are ill-defined after merging, we plot the widest accessible pre-merging range. The same analysis for structure~B is shown in \autoref{figS:merging_tracking}(f--j).

Structure~A exhibits weak momentum anisotropy ($F_k \approx 1$) but appreciable frequency anisotropy, with $F_f \rightarrow 1$ as $L$ is tuned, indicating that the M- and X-BICs merge at $\Gamma$ nearly simultaneously. Structure~B shows the same qualitative behavior, albeit with pronounced anisotropy in both momentum and frequency. These results confirm that both structures belong to the anisotropic-merging regime. For the C and D families discussed in the main text, the cross-merging behavior is already evident and is therefore not repeated here.

To demonstrate that anisotropic merging and cross-merging are continuously connected, we introduce structure~F with $(P,t,r)=(450~\mathrm{nm},\,200~\mathrm{nm},\,0.3P)$ and $L$ tuned around $250~\mathrm{nm}$. Although structure~F initially resembles anisotropic merging, a closer inspection reveals a symmetry-direction transfer characteristic of cross-merging. As shown in \autoref{figS:merging_tracking}(k), the M-BIC approaches $\Gamma$, crosses through it, and converts into a second X-BIC, while the original X-BIC has not yet merged. The two X-direction BICs subsequently merge at a finite off-$\Gamma$ wave vector along $\Gamma$--X. This evolution identifies structure~F as an intermediate scenario, in which an anisotropic-merging-like configuration develops into cross-merging. Consistently, the anisotropic merging factors exhibit the singular behavior characteristic of cross-merging [\autoref{figS:merging_tracking}(o)].

\begin{figure}[H]
    \centering
    \includegraphics[width=0.85\linewidth]{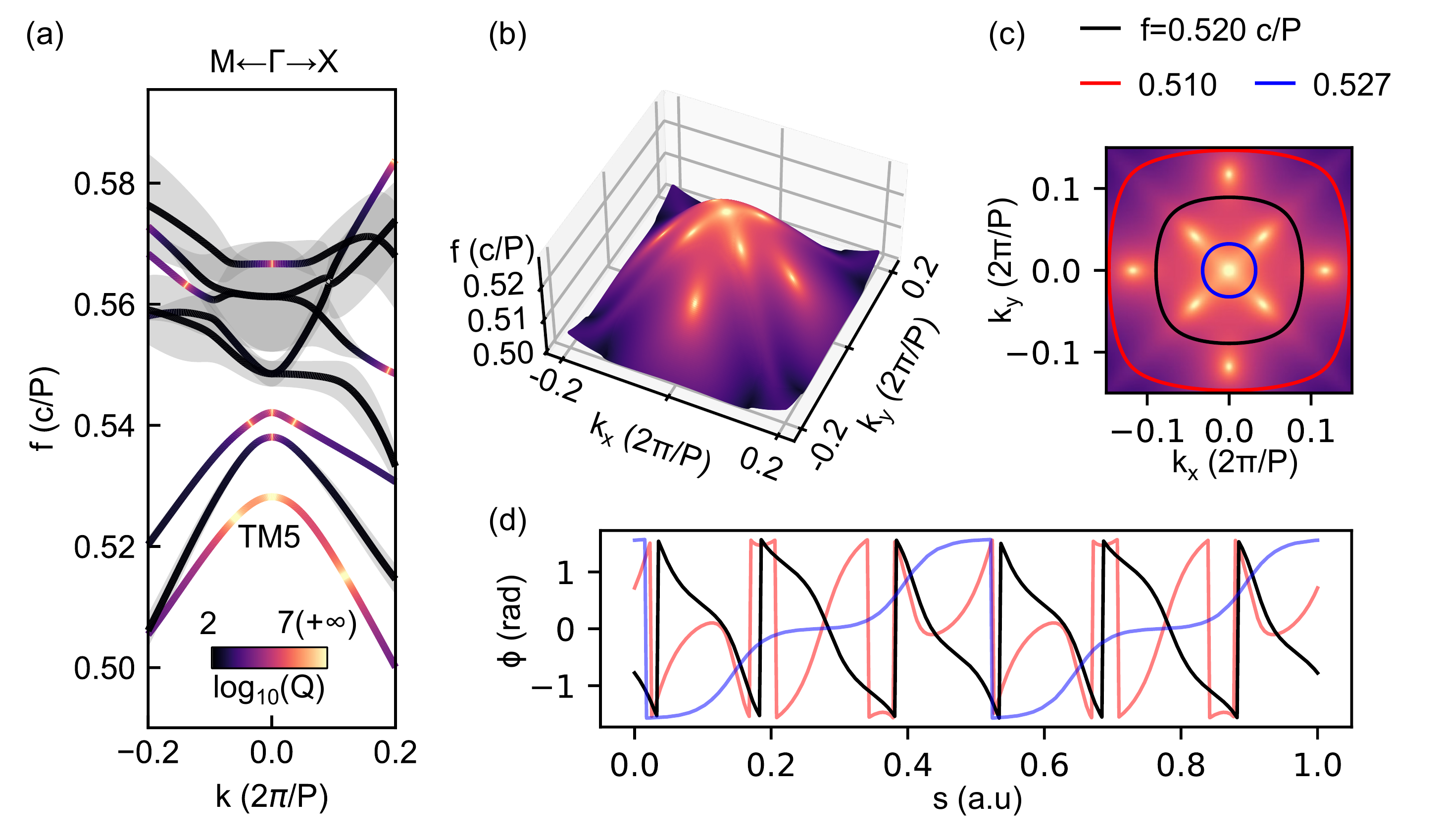}
    \caption{
    Momentum-space dispersion and BIC distribution for structure~F at $L=250~\mathrm{nm}$, shown in a style analogous to Fig.~1 of the main text.
    }
    \label{figS:strF_details}
\end{figure}

For completeness, \autoref{figS:strF_details} presents the momentum-space band structure of structure~F at $L=250~\mathrm{nm}$ in the same style as Fig.~1 of the main text. An annular high-$Q$ intersection appears on the TM\textsubscript{5} band, with unequal M- and X-BIC frequencies. Although the TM\textsubscript{5} guided resonance is only weakly anisotropic, the strongly anisotropic FP interference drives the intermediate merging behavior. The resulting iso-frequency contours remain relatively circular, which is advantageous for high-quality vortex-beam generation.

\section{Polarization-Resolved Anisotropic Dispersion in Square-Hole Photonic Crystal Slabs}

In the main text, the square-hole photonic crystal slab (structure~B) is used as a representative platform to realize pronounced anisotropic merging. Here we provide a more detailed, polarization-resolved analysis to explicitly demonstrate that this behavior originates from the anisotropic FP interference, rather than from the intrinsic dispersion of the GR. As shown in \autoref{figS:square_hole}(a), structure~B uses $(P,t,d)=(400~\mathrm{nm},\,200~\mathrm{nm},\,0.6P)$, and the cavity thickness is tuned across three representative cases, namely $L=170~\mathrm{nm}$ (B$_1$), $L=180~\mathrm{nm}$ (B$_2$), and $L=190~\mathrm{nm}$ (B$_3$). The corresponding momentum-space band structures are plotted in \autoref{figS:square_hole}(c), (d), and (g).

\begin{figure}[H]
    \centering
    \includegraphics[width=0.85\linewidth]{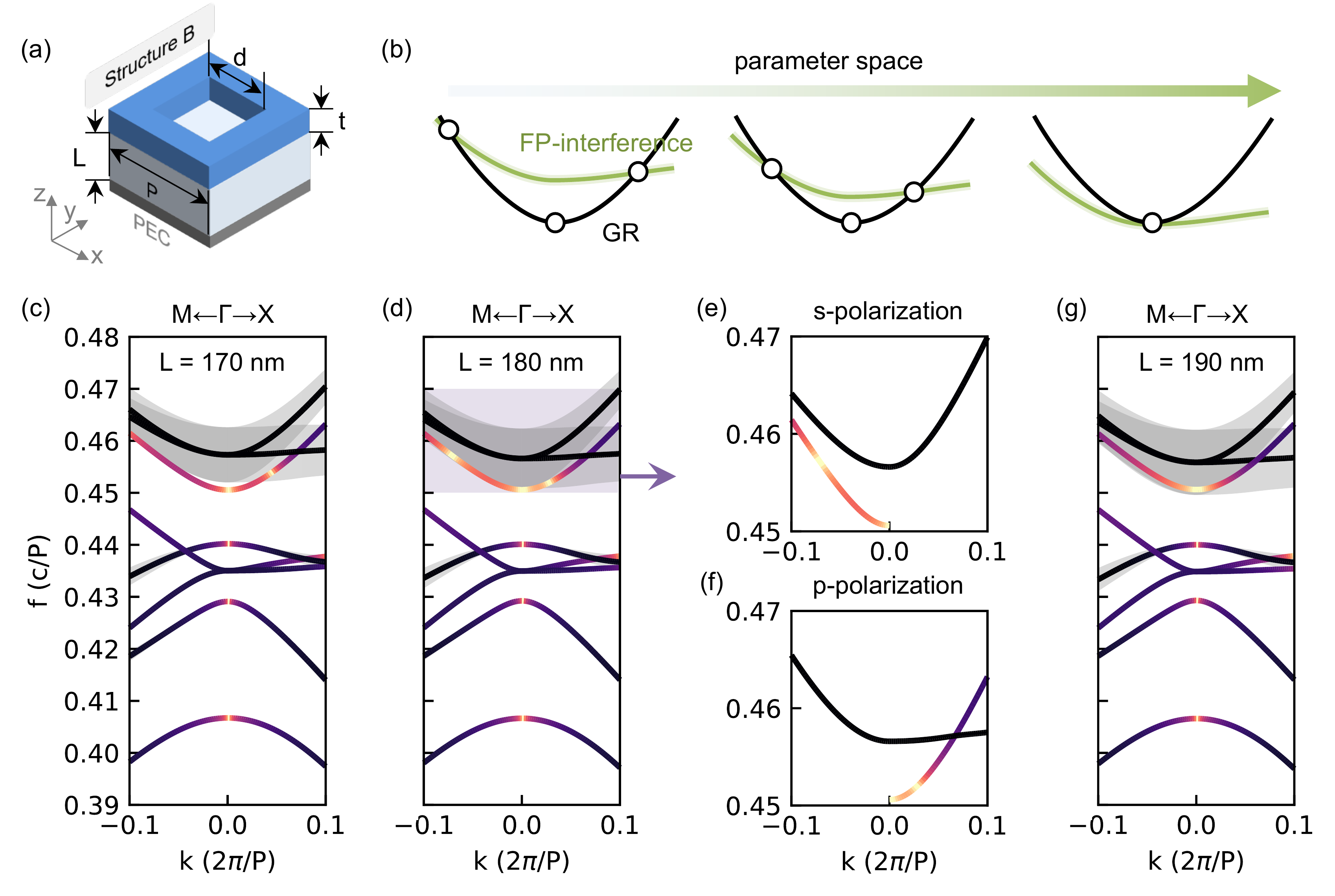}
    \caption{
    \textbf{Anisotropic merging in the square-hole PCS.}
    (a) Schematic of the square-hole PCS–FP cavity.
    (b) Geometric model illustrating anisotropic merging driven by anisotropic FP interference.
    (c,d,g) 1D dispersions of structures B\textsubscript{1}, B\textsubscript{2}, and B\textsubscript{3} along the M–$\Gamma$–X directions.
    (e,f) $s$- and $p$-polarized FP-interference and GR dispersions, respectively, demonstrating the anisotropic FP–GR intersection pattern shown in (b).
    }
    \label{figS:square_hole}
\end{figure}

Because FW-/FP-type BICs originate from destructive interference within the same radiation channel, it is instructive to resolve the system into $s$- and $p$-polarized components. The polarization-resolved FP and GR dispersions are shown in \autoref{figS:square_hole}(e,f). In this representation, the GR-BIC associated with the GR branch is clearly identified as carrying a topological charge of $-1$. By matching the polarization channels of the FP and GR modes, we observe a strongly anisotropic FP dispersion.

As depicted in \autoref{figS:square_hole}(b), the FP dispersion is significantly flatter along the $\Gamma$–X direction than along $\Gamma$–M. Consequently, the off-$\Gamma$ BICs located on $\Gamma$–X appear at lower frequencies compared with those on $\Gamma$–M. Under variation of $L$, these BICs move in frequency but eventually merge simultaneously at $\Gamma$, forming an anisotropic merging process consistent with the mechanisms described in the main text.

These results provide a second, clear verification that anisotropic merging arises directly from the anisotropic dispersion of the underlying momentum-space modes.

\section{Detailed evolution of states of polarization and eigenfields in the main text}

\begin{figure}[h]
    \centering
    \includegraphics[width=1\linewidth]{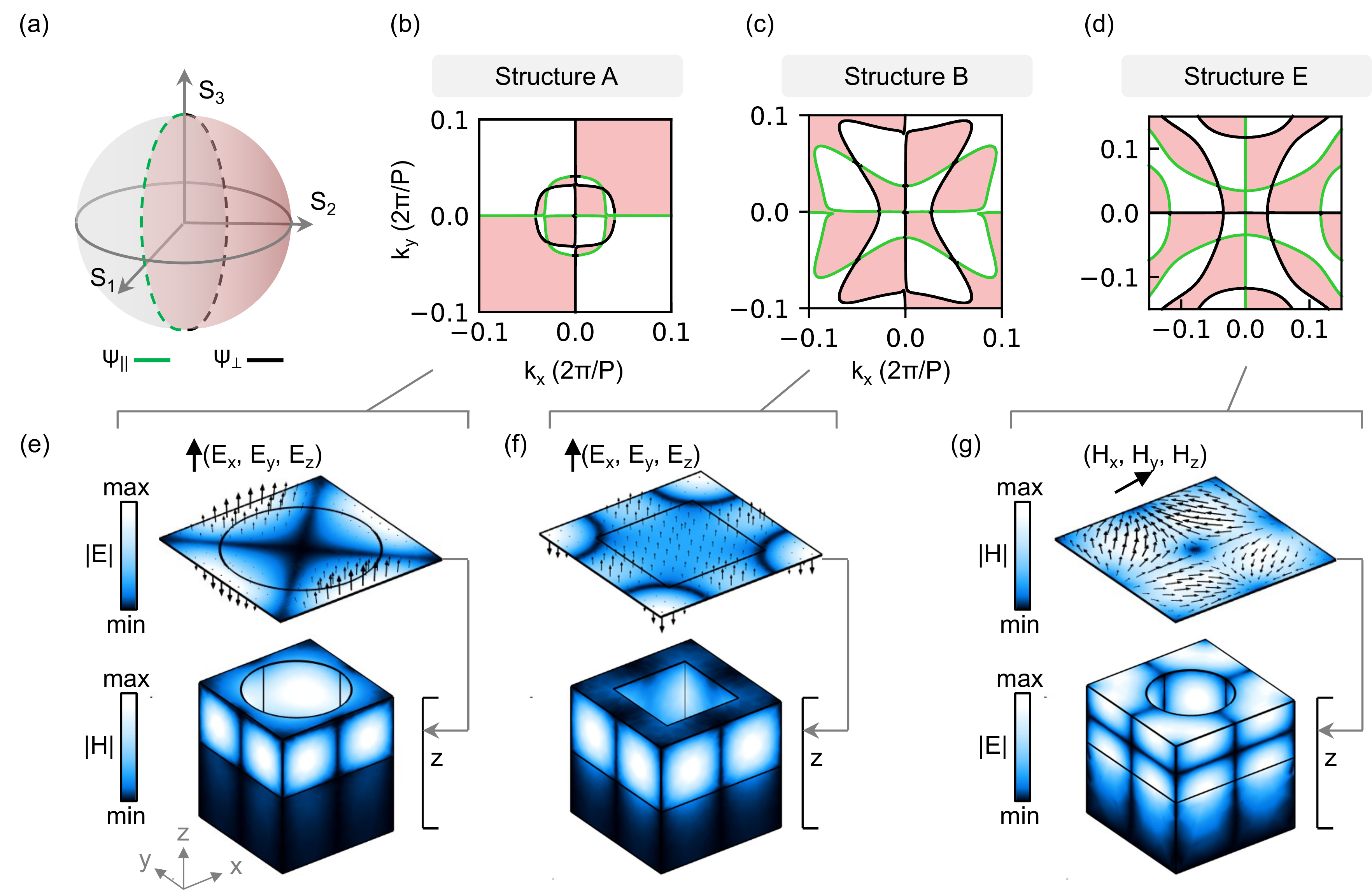}
    \caption{
        \textbf{SOPs for structures A, B and E in the main text.}
        (a) Schematic diagram of a polarization graph and its dual graph. In the polarization graph, the edges correspond to polarization nodal lines of the major axis of the polarization ellipse. Green lines indicate nodal lines of the major-axis angle $\psi_k$, whereas black lines indicate $\psi_\perp = \psi_k + \pi/2$. Bound states in the continuum (BICs) and C points correspond to the intersection and joint points of edges with different colors, respectively. The dual graph is indicated by the purple vertices and edges.
        (b,c,d) Polarization graphs for structures A, B and E, respectively. (e,f,g) $\Gamma$-point eigenfields for  structures A, B and E, respectively.
    }
    \label{fig:SOPs+NearField_strABE}
\end{figure}

\begin{figure}[h]
    \centering
    \includegraphics[width=1\linewidth]{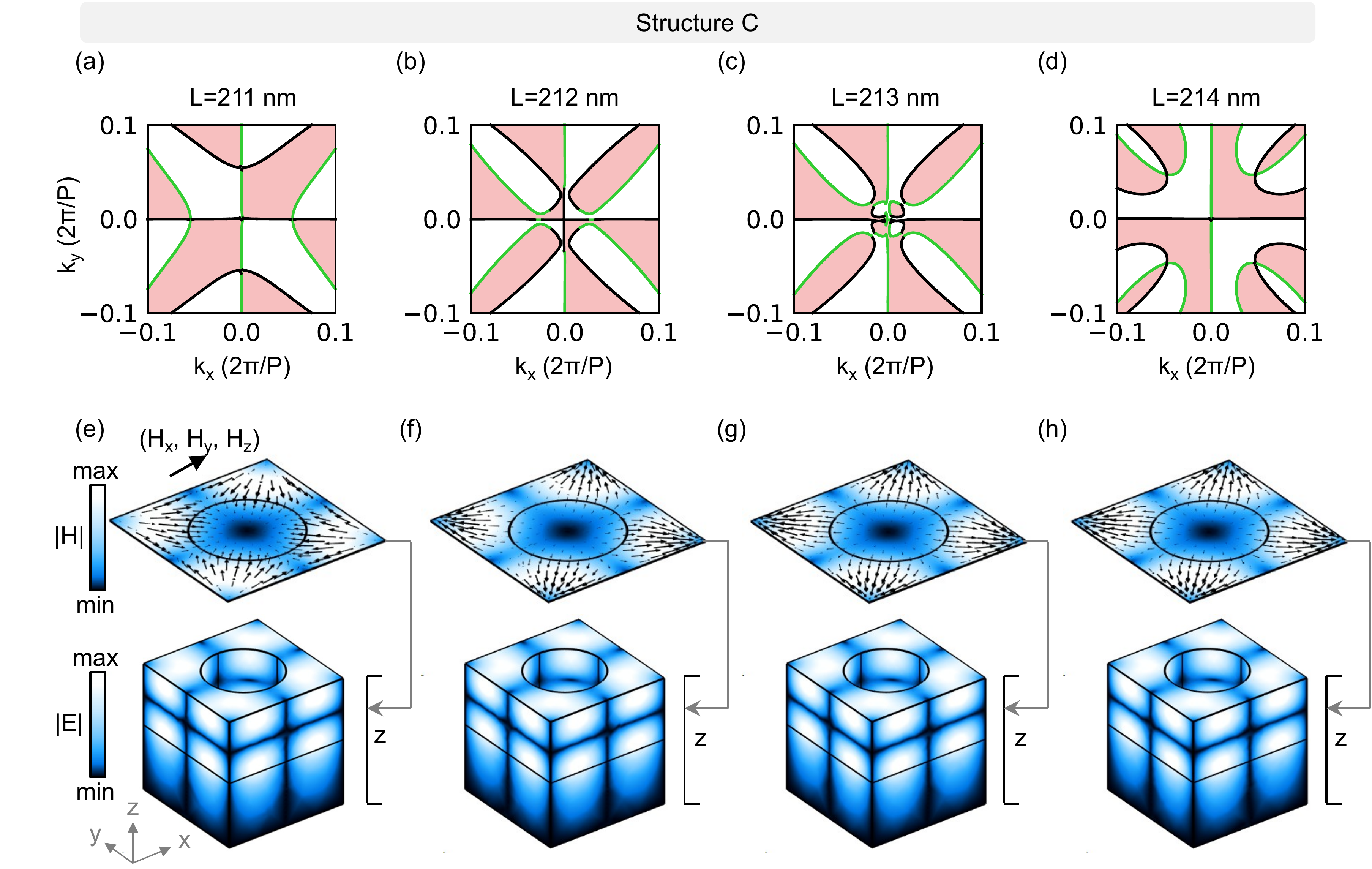}
    \caption{
        \textbf{SOPs for structures C\textsubscript{1}, C\textsubscript{2}, C\textsubscript{3} in the main text, and an additional structure with $L = 213~\mathrm{nm}$.}
      (a-d)  Polarization graphs showing the evolution of BIC positions and C points as the FP thickness $L$ is tuned across the cross-merging regime. (e-h) Corresponded $\Gamma$-point eigenfields.
    }
    \label{fig:SOPs+NearField_strC1243}
\end{figure}

\begin{figure}[h]
    \centering
    \includegraphics[width=0.75\linewidth]{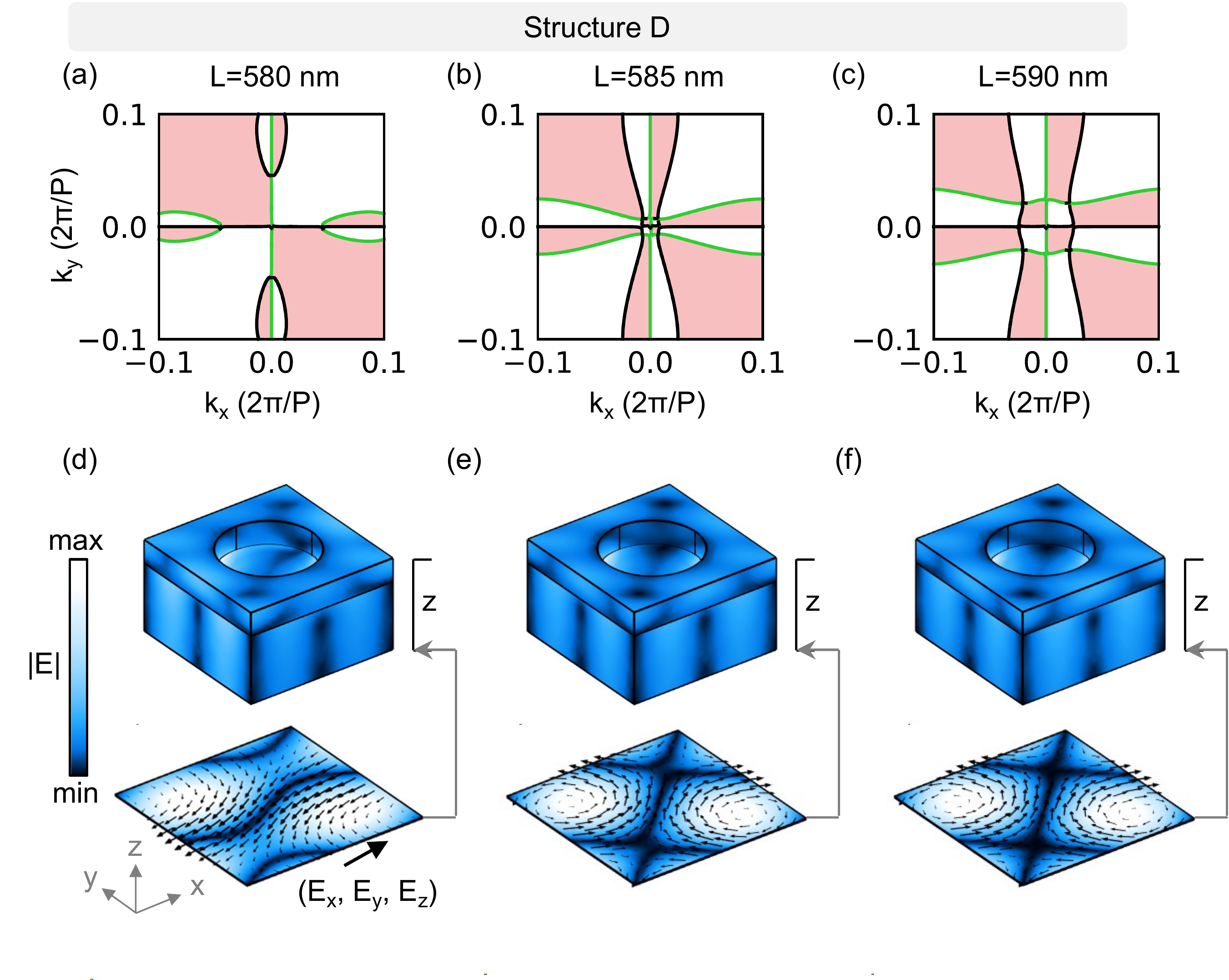}
    \caption{
        \textbf{SOPs for structures D\textsubscript{1}, D\textsubscript{2}, and D\textsubscript{3} in the main text.}
      (a-c)  Polarization graphs for the lower branch near the degeneracy point, illustrating the motion and cross-merging of off-$\Gamma$ BICs under variation of the FP thickness $L$. (d-f) Corresponded $\Gamma$-point eigenfields.
    }
    \label{fig:SOPs+NearField_strD123}
\end{figure}

\begin{figure}[h]
    \centering
    \includegraphics[width=0.6\linewidth]{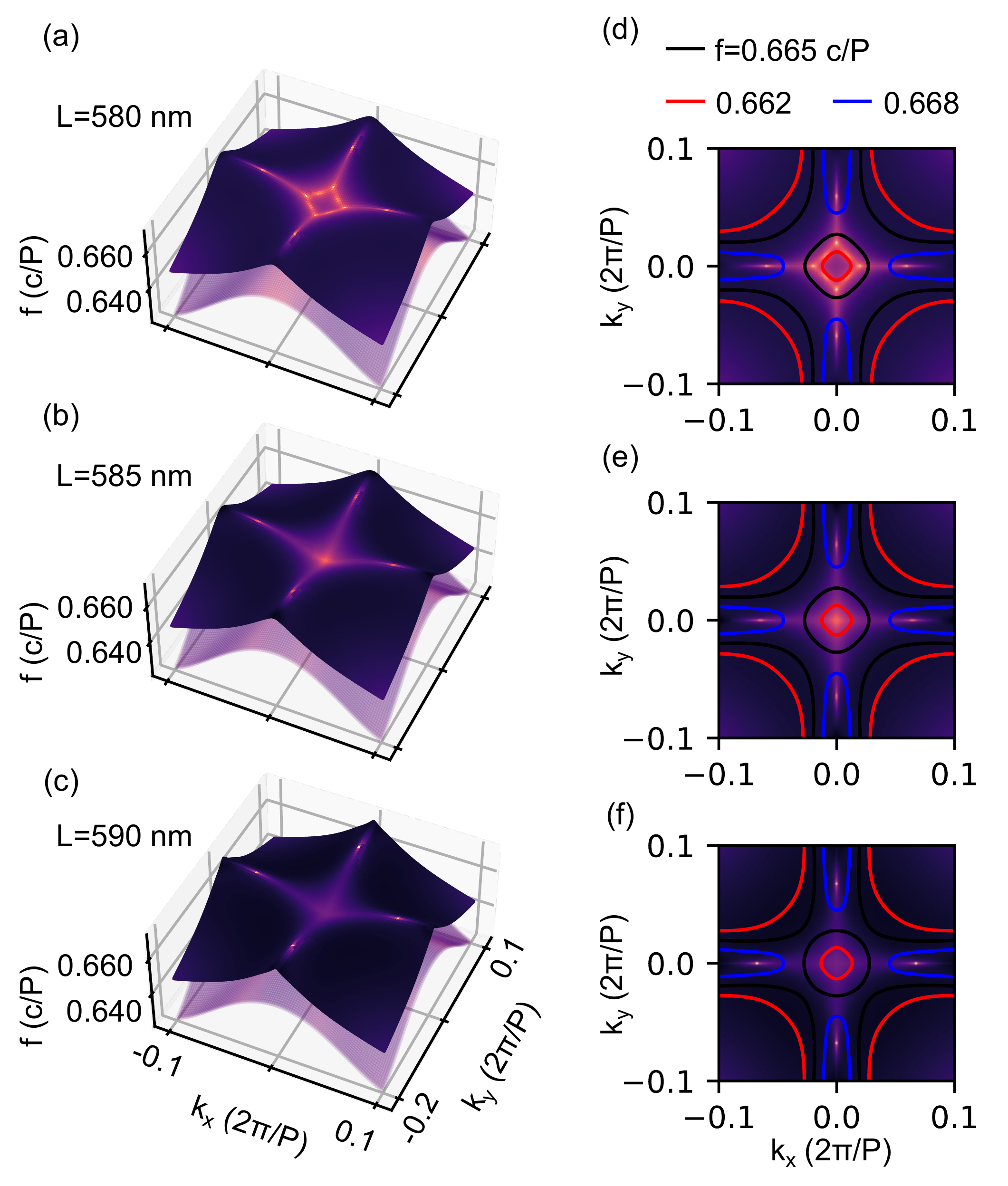}
    \caption{
        \textbf{Cross-merging at a degeneracy point (DP) for the upper branch.}
        (a-c) 2D $Q$-colored dispersions of the upper branch for structures D\textsubscript{1}, D\textsubscript{2}, and D\textsubscript{3} with different FP thicknesses $L$.
        (d-f) Representative iso-frequency contours (solid lines) overlaid on the corresponding $Q$ maps for D\textsubscript{1}, D\textsubscript{2}, D\textsubscript{3}.
    }
    \label{fig:upperDP}
\end{figure}

In this section we provide the full momentum-space states of polarization (SOPs) for all structures discussed in the main text, namely A, B, E, C\textsubscript{1}, C\textsubscript{2}, C\textsubscript{3}, D\textsubscript{1}, D\textsubscript{2}, and D\textsubscript{3}. The corresponding polarization graphs and eigenfields are shown in \autoref{fig:SOPs+NearField_strABE}, \autoref{fig:SOPs+NearField_strC1243}, and \autoref{fig:SOPs+NearField_strD123}. For the C-family, we also include an additional structure with FP thickness $L=213~\mathrm{nm}$ to illustrate the fine evolution of the merging process. For the D-family, we further analyze the BIC merging on the upper branch at the degeneracy point (DP), as shown in \autoref{fig:upperDP}.

Following Ref.~\cite{RN247}, the SOPs are visualized using polarization graphs defined on the Poincar\'e sphere representation of the far-field polarization. The basic construction is summarized in \autoref{fig:SOPs+NearField_strABE}(a).

For the eigenfield analysis, we plot the eigenfields of the target $\Gamma$-point mode for each structure. Both electric- and magnetic-field amplitudes are shown, together with field arrows (see figure captions for details). In addition to the 3D surface field distributions, we also display field distributions on selected cross sections. In \autoref{fig:SOPs+NearField_strABE} and \autoref{fig:SOPs+NearField_strC1243}, the fields are shown at the mid-plane of the PCS slab, whereas in \autoref{fig:SOPs+NearField_strD123} the fields are plotted at the bottom Perfect Electric Conductor (PEC) plane, where $H_z = E_x = E_y = 0$.

In \autoref{fig:SOPs+NearField_strABE}(e,f,g), the field profiles clearly exhibit the $C_4$ symmetry characteristic of symmetry-protected $\Gamma$-point BICs, consistent with their decoupling from free-space plane waves\cite{RN194}. One natural question is whether the merging of off-$\Gamma$ BICs alters the eigenfield patterns—particularly in the case of cross merging, which yields a BIC with high-order topological charge at the $\Gamma$ point. To address this, we examine \autoref{fig:SOPs+NearField_strC1243} and \autoref{fig:SOPs+NearField_strD123}, which span the entire cross-merging process from left to right. Notably, the eigenfield patterns remain essentially unchanged throughout. This behavior is expected: unlike symmetry-protected BICs, the BIC with high-order TCs obtained through cross merging are not robust under parameter variations and therefore do not enforce symmetry-protected field profiles. This observation is fully consistent with the discussion in the main text.

Finally, we present the detailed evolution of BICs on the upper branch at the DP for structures D\textsubscript{1}, D\textsubscript{2}, and D\textsubscript{3}, as shown in \autoref{fig:upperDP}. These plots complement the main-text results for the lower branch and demonstrate that the cross-merging mechanism and high-order TC formation also occur on the upper band.

\newpage


In all polarization graphs shown above, the SOP at each Bloch wave vector $\mathbf{k}$ is characterized by the polarization angle $\phi(\mathbf{k})$ of the major axis of the polarization ellipse, defined modulo $\pi$. Numerically, we evaluate $\phi(\mathbf{k})$ on a regular $\mathbf{k}$-space grid from the complex far-field amplitudes and wrap it into the interval $[0,\pi)$.

The nodal lines of the major axis are then extracted from the condition $\sin(2\phi)=0$ using a marching-squares algorithm on this grid, which yields a set of polygonal lines approximating the zero-level set of $\sin(2\phi)$. Along each such line, we inspect the sign of $\cos(2\phi)$ on the two sides of the nodal line: if $\cos(2\phi)$ is predominantly positive, the segment is assigned to the family with $\phi\approx 0,\pi$; if it is predominantly negative, the segment is assigned to the family with $\phi\approx \pi/2$. Segments for which both sides are nearly zero are marked as uncertain and plotted separately. By stitching together segments with consistent classification, we obtain two families of nodal lines, which form the colored edges of the polarization graph used in \autoref{fig:SOPs+NearField_strABE}–\autoref{fig:SOPs+NearField_strD123}. Intersection points of edges from different families correspond to C points, and their global connectivity encodes the underlying topological charges of the BICs and the surrounding polarization texture.

\clearpage

\section{Summary of Reported $\Gamma$-Point Topological Charges in Finite-Symmetry Photonic Structures}

In this section, we compile the momentum-space far-field TCs at the $\Gamma$ point that have been reported to date in photonic crystal slabs (PCSs) and quasicrystalline PCSs with finite rotational symmetries. The summary in \autoref{tab:GammaTC} includes both numerical and experimental results from the literature, covering isolated BICs, degenerate points (DPs), and special doubly degenerate BICs (D-BICs) in $C_6$ systems. For higher-order quasicrystalline symmetries ($C_8$, $C_{10}$, $C_{12}$, ...), conventional BIC descriptions become inaccurate due to the absence of strict Bloch periodicity; nevertheless, their reported momentum-space winding numbers are included for completeness.

In some references\cite{RN694,RN831}, the TC of a DP at the $\Gamma$ point is assigned as zero, due to its strict polarization independence under orthogonal degeneracy. Here, however, we emphasize that the winding number of the momentum-space polarization field remains well defined when one examines either the upper or the lower branch individually. Moreover, DPs can also generate vortex beams through the Pancharatnam–Berry (PB) phase mechanism\cite{RN631,RN436}. Therefore, in this summary we adopt a nonzero TC notation for DPs, consistent with the definition used for other BIC-related topological charges.

\begin{table}[H]
\centering
\begin{tabular}{|c|c|c|c|}
\hline
\textbf{Symmetry} & \textbf{Type} & \textbf{Reference} & \textbf{TC} \\
\hline
\multirow{2}{*}{$C_2$}
 & \multirow{2}{*}{BIC}
 & \cite{RN475,RN146} & $+1$ \\ \cline{3-4}
 & & \cite{RN475,RN146} & $-1$ \\
\hline
\multirow{1}{*}{$C_3$}
 & \multirow{1}{*}{BIC}
 & \cite{RN246} & $+1$ \\
\hline
\multirow{4}{*}{$C_4$}
 & \multirow{2}{*}{BIC}
 & \cite{RN240} & $+1$ \\ \cline{3-4}
 & & \cite{RN475} & $-1$ \\ \cline{2-4}
 & \multirow{2}{*}{DP}
 & \cite{RN317,Li2025DegenerateBIC} & $-1$ \\ \cline{3-4}
 & & \cite{wang2025} & $+1$ \\
\hline
\multirow{4}{*}{$C_6$}
 & \multirow{2}{*}{BIC}
 & \cite{RN831} & $+1$ \\ \cline{3-4}
 & & \cite{RN831,RN246,RN698,RN695} & $-2$ \\ \cline{2-4}
 & DP
 & \cite{RN695} & $+1$ \\ \cline{2-4}
 & D-BIC
 & \cite{RN831} & $-1$ \\
\hline
$C_8$    & \textemdash & \cite{RN322,RN836} & $-3$ \\ \hline
$C_{10}$ & \textemdash & \cite{RN836}        & $-4$ \\ \hline
$C_{12}$ & \textemdash & \cite{RN836}        & $-5$ \\ \hline
\textcolor{red}{$C_4$}  & \textcolor{red}{BIC \& DP}      & \textcolor{red}{This work}    & \textcolor{red}{$\pm 3$} \\ \hline
\end{tabular}

\caption{Reported $\Gamma$-point topological charges in (quasi) PCSs with finite rotational symmetries. \small{``BIC'': isolated nondegenerate BIC; ``DP'': orthogonally degenerate states with finite radiative $Q$; ``D-BIC'': doubly degenerate BIC specific to $C_6$ systems; ``---'': quasicrystal regimes where conventional BIC descriptions become inaccurate.}}
\label{tab:GammaTC}
\end{table}

\newpage

\section{Generation of phase vortices in momentum space}

In this section we detail how the high-order phase vortices in momentum space, shown in Figure 5 of the main text, are obtained from single GR modes associated with BICs, and how they are subsequently propagated into real space using an angular-spectrum method. We first recall the basic single-mode picture for vortex generation, then describe two complementary numerical approaches: a direct scattering simulation in COMSOL by scanning $(k_x,k_y)$, and an eigenmode-based interpolation scheme that enables efficient, high-precision evaluation of the cross-polarized response even for ultra-high-$Q$ resonances. Finally, we summarize the angular-spectrum propagation used to obtain real-space vortex beams and their non-diffracting behavior.

\subsection*{Single-mode picture for BIC-enabled vortex generation}

As discussed in Refs.~\cite{RN269,RN400} and in the main text, the GR of a PCS can be modeled as an anisotropic, momentum-dependent wave plate acting on the incident polarization state. In local coordinates aligned with the resonance polarization at azimuthal angle $\phi(\mathbf{k})$, the Jones matrix takes the form
\begin{equation}
\mathbf{J}_{\mathrm{local}}(\mathbf{k},\omega) =
\begin{bmatrix}
s_{\mathrm{res-x}}(\mathbf{k},\omega) + s(\omega) & 0 \\
0 & s_{\mathrm{res-y}}(\mathbf{k},\omega) + s(\omega)
\end{bmatrix},
\end{equation}
where $s(\omega)$ is the non-resonant background response, and $s_{\mathrm{res-x}}$, $s_{\mathrm{res-y}}$ describe the resonant contributions for two orthogonal linear polarizations. For a single, linearly polarized GR, one may take $s_{\mathrm{res-y}}\approx 0$ and write $s_{\mathrm{res-x}}\equiv s_{\mathrm{res}}(\mathbf{k},\omega)$. The polarization dependence is then quantified by $s_{\mathrm{res}}(\mathbf{k},\omega)$.

After rotating back to the global $(x,y)$ basis and transforming to the circular basis $\{\ket{R},\ket{L}\}$, the scattered field under right-circular input $\ket{R}$ reads~\cite{RN269}
\begin{equation}
\ket{E_{\mathrm{out}}(\mathbf{k},\omega)}
= \frac{1}{2}\big[s_{\mathrm{res}}(\mathbf{k},\omega)
 + 2s(\omega)\big]\ket{R}
+ \frac{1}{2} s_{\mathrm{res}}(\mathbf{k},\omega)e^{-2i\phi(\mathbf{k})}\ket{L}.
\label{eq:PB-cross}
\end{equation}
Thus, the cross-polarized component is
$s_{\mathrm{cross}}(\mathbf{k},\omega)
\equiv \braket{L|E_{\mathrm{out}}}
= \frac{1}{2} s_{\mathrm{res}}(\mathbf{k},\omega)e^{-2i\phi(\mathbf{k})},$
which carries a Pancharatnam–Berry phase $\Phi_{\mathrm{PB}}(\mathbf{k})=-2\phi(\mathbf{k})$ through the factor $e^{-2i\phi(\mathbf{k})}$. For a GR whose major-axis angle winds as $\phi(\theta_k)=q\,\theta_k$ along the iso-frequency contour (with $\theta_k$ the azimuthal angle in momentum space and $q$ the polarization topological charge), this cross-polarized channel acquires a vortex phase
$s_{\mathrm{cross}}(\theta_k,\omega)\propto s_{\mathrm{res}}(\theta_k,\omega)
\,e^{-2iq\theta_k},$
corresponding to $\mathrm{OAM}=2q$ in the reflected or transmitted cross-polarized beam. In a $C_4$-symmetric PCS, such polarization vortices and their winding number $q$ are enforced by symmetry and topology~\cite{wang2025}, while the BIC (or near-BIC) condition controls the strength and spectral width of $s_{\mathrm{res}}$ via the resonance $Q$ factor.

For a single linear-polarized resonance at frequency $\omega_0(\mathbf{k})$, temporal coupled-mode theory (TCMT) gives a Lorentzian dependence~\cite{RN239}
\begin{equation}
s_{\mathrm{res}}(\mathbf{k},\omega)
= -\frac{d_s^2(\mathbf{k})+d_p^2(\mathbf{k})}
{\gamma_{\mathrm{rad}}(\mathbf{k})+\gamma_{\mathrm{abs}}-i[\omega-\omega_0(\mathbf{k})]},
\end{equation}
where $\gamma_{\mathrm{rad}}$ and $\gamma_{\mathrm{abs}}$ are the radiative and absorptive decay rates, and $d_{s,p}$ are the coupling amplitudes to $s$- and $p$-polarized channels (in our case, $d_s^2(\mathbf{k})+d_p^2(\mathbf{k})$ $=2\gamma_{\mathrm{rad}}(\mathbf{k})$ for reflective metasurface). Near a BIC, $\gamma_{\mathrm{rad}}(\mathbf{k})\to 0$ and the resonance becomes extremely narrow, which enhances the peak cross-polarization but also makes direct frequency and momentum sampling numerically demanding.

\subsection*{Direct COMSOL simulation by $(k_x,k_y)$ scanning}

A conceptually straightforward way to obtain $s_{\mathrm{cross}}(\mathbf{k},\omega)$ is to perform direct frequency- and angle-resolved scattering simulations in COMSOL. For a fixed normalized frequency $f=\omega P/(2\pi c)$, we proceed as follows:

\begin{enumerate}
    \item Fix the operating frequency $f$ and choose a rectangular sampling grid in momentum space, $(k_x,k_y)$, within the first Brillouin zone. In practice we parameterize $k_x$ and $k_y$ in units of $2\pi/P$.
    \item For each $(k_x,k_y)$ point, set the incident plane wave with right- (or left-) circular polarization and in-plane wave vector $(k_x,k_y)$, and compute the outgoing fields.
    \item Project the reflected (or transmitted) field onto the cross-circular channel $\ket{L}$ to extract the complex cross-polarization coefficient $s_{\mathrm{cross}}(k_x,k_y)$.
\end{enumerate}

By assembling $s_{\mathrm{cross}}(k_x,k_y)$ over the grid, one directly obtains the momentum-space vortex pattern, whose phase singularities and azimuthal phase winding encode the OAM. This procedure is numerically robust and non-perturbative, but for ultra-high-$Q$ modes near a BIC, the spectral response becomes extremely sharp in both frequency and momentum space. In such cases, resolving the narrow resonance in COMSOL requires a very fine sampling in $(k_x,k_y)$ and careful frequency tuning, leading to prohibitive computation times.

\subsection*{Eigenmode-based interpolation and single-mode approximation}

To efficiently handle high-$Q$ modes and exploit the underlying GR topology, we adopt an eigenmode-based interpolation method that reconstructs $s_{\mathrm{cross}}(\mathbf{k},\omega)$ from the band structure and eigen-polarization data, under a single-linear-mode approximation. This approximation requires only three quantities from the target band: the real part of the eigenfrequency $\omega_n'(\mathbf{k})$, the radiative linewidth $\gamma_n(\mathbf{k})$, and the far-field polarization angle $\phi_n(\mathbf{k})$.

We begin by computing the complex eigenfrequencies $\omega_n(\mathbf{k})=\omega_n'(\mathbf{k})-i\gamma_n(\mathbf{k})$ and the major-axis polarization angle $\phi_n(\mathbf{k})$ for the relevant band~$n$ (e.g., TM\textsubscript{5} for structure~B or the lower/upper branch near the DP for structure~D) on a coarse $\mathbf{k}$-space grid. Because the circular-polarization fraction of the target modes is extremely small and no parasitic resonances lie nearby, treating each band as a single linearly polarized GR mode is well justified.

The quantities $\omega_n'(\mathbf{k})$, $\gamma_n(\mathbf{k})$, and $\phi_n(\mathbf{k})$ are then interpolated onto a finer momentum grid using spline or higher-order interpolation. For a given operating frequency $\omega$, the cross-polarized response is modeled as
\begin{equation}
s_{\mathrm{cross}}^{(n)}(\mathbf{k},\omega)
\simeq \exp[-2i\phi_n(\mathbf{k})]\,
\frac{\gamma_n(\mathbf{k})}{\gamma_n(\mathbf{k}) - i[\omega-\omega_n'(\mathbf{k})]},
\label{eq:single-mode-cross}
\end{equation}
where the Lorentzian factor captures the resonance detuning and linewidth, and the azimuthal phase $\exp[-2i\phi_n(\mathbf{k})]$ encodes the underlying polarization vortex of the GR.

In practice, the numerical procedure is:
\begin{enumerate}
    \item Load the eigenmode data for the desired structure (e.g., structure B with parameters $(P,t,L,r)$) and precompute $\omega_n(\mathbf{k})$ and $\phi_n(\mathbf{k})$ on a coarse grid.
    \item Interpolate the data to a finer momentum grid by a user-chosen refinement factor (e.g., by a factor of 5 in each direction) to improve spatial resolution of the vortex pattern.
    \item For a fixed target frequency $f=\omega P/(2\pi c)$, evaluate $s_{\mathrm{cross}}^{(n)}(\mathbf{k},\omega)$ using a Lorentzian line shape centered at $\omega_n'(\mathbf{k})$ with width $\gamma_n(\mathbf{k})$, multiplied by the phase factor $\exp[-2i\phi_n(\mathbf{k})]$ as in \eqref{eq:single-mode-cross}.
\end{enumerate}

In code, because the resonance is treated analytically in frequency and only the geometric quantities $\omega_n'(\mathbf{k})$, $\gamma_n(\mathbf{k})$, and $\phi_n(\mathbf{k})$ are interpolated, this approach can achieve effectively arbitrary precision in the cross-polarization map, even when the underlying band data are computed on a finite-resolution grid. It also avoids repeated full-wave scattering simulations for each $(k_x,k_y)$ point.

This single-mode interpolation method is used to generate the momentum-space vortex patterns shown in Figure 5(c) and Figure 5(d) of the main text, where structure B yields a nearly circular vortex in $s_{\mathrm{cross}}(\mathbf{k})$ at $f=0.453\,c/P$, corresponding to $|{\rm OAM}|=6$ with improved beam quality.

\subsection*{Angular-spectrum propagation and non-diffracting beams}

Once the complex cross-polarized spectrum $s_{\mathrm{cross}}(k_x,k_y)$ is obtained (either by direct COMSOL scanning or by eigenmode interpolation), we treat it as the initial angular spectrum $U_0(k_x,k_y)$ of the reflected field at the PCS plane. The subsequent propagation into real space is computed using the standard angular-spectrum method.

We first regard $(k_x,k_y)$ as the in-plane spatial frequencies of the field at $z=0$, and write the field at that plane as
\begin{equation}
E(x,y;0) = \iint U_0(k_x,k_y)\,e^{i(k_x x + k_y y)}\,\mathrm{d}k_x\,\mathrm{d}k_y.
\end{equation}
In free space, each plane-wave component acquires a longitudinal phase factor
\begin{equation}
H(k_x,k_y;z) = \exp\big[i k_z(k_x,k_y)\,z\big],
\end{equation}
where $k_z(k_x,k_y) = \sqrt{k_0^2 - k_x^2 - k_y^2}$ and $k_0$ is the free-space wavenumber at the operating frequency. The field at a propagation distance $z$ is then given by
\begin{equation}
E(x,y;z) = \iint U_0(k_x,k_y)\,H(k_x,k_y;z)\,e^{i(k_x x + k_y y)}\,\mathrm{d}k_x\,\mathrm{d}k_y,
\end{equation}
which is evaluated numerically using FFT-based routines by interpreting $U_0(k_x,k_y)$ as the discrete Fourier spectrum of the field.

By repeating this calculation for a set of propagation distances $z$, one obtains both the transverse intensity profiles $|E(x,y;z)|^2$ and longitudinal cuts such as $|E(x,z)|^2$ at fixed $y=0$. This procedure produces the real-space phase vortices and non-diffracting beams shown in Figure 5(b–e) of the main text. Near a merging BIC, the large radiative $Q$ factor effectively narrows the momentum-space bandwidth of $U_0(k_x,k_y)$, leading to slowly varying, extended propagation along $z$—a hallmark of non-diffracting behavior~\cite{Kim2025, RN544}. Moreover, the sign of the group velocity of the underlying GR mode sets a preferred propagation direction; in our examples the beams propagate non-diffractively co-directed with the negative group-velocity branch, consistent with the theoretical prediction in ref~\cite{Kim2025}.

\bibliography{ref}

@article{RN1041,
   author = {Alagappan, G. and Garcia-Vidal, F. J. and Png, C. E.},
   title = {Fabry-Perot Resonances in Bilayer Metasurfaces},
   journal = {Phys Rev Lett},
   volume = {133},
   number = {22},
   pages = {226901},
   ISSN = {1079-7114 (Electronic)
0031-9007 (Linking)},
   DOI = {10.1103/PhysRevLett.133.226901},
   url = {https://www.ncbi.nlm.nih.gov/pubmed/39672113},
   year = {2024},
   type = {Journal Article}
}

@article{RN722,
  title = {Formation of Bound States in the Continuum in Hybrid Plasmonic-Photonic Systems},
  author = {Azzam, Shaimaa I. and Shalaev, Vladimir M. and Boltasseva, Alexandra and Kildishev, Alexander V.},
  journal = {Phys. Rev. Lett.},
  volume = {121},
  issue = {25},
  pages = {253901},
  numpages = {6},
  year = {2018},
  month = {Dec},
  publisher = {American Physical Society},
  doi = {10.1103/PhysRevLett.121.253901},
  url = {https://link.aps.org/doi/10.1103/PhysRevLett.121.253901}
}

@article{RN818,
author = {Andrey A. Bogdanov and Kirill L. Koshelev and Polina V. Kapitanova and Mikhail V. Rybin and Sergey A. Gladyshev and Zarina F. Sadrieva and Kirill B. Samusev and Yuri S. Kivshar and Mikhail F. Limonov},
title = {{Bound states in the continuum and Fano resonances in the strong mode coupling regime}},
volume = {1},
journal = {Advanced Photonics},
number = {1},
publisher = {SPIE},
pages = {016001},
year = {2019},
doi = {10.1117/1.AP.1.1.016001},
URL = {https://doi.org/10.1117/1.AP.1.1.016001}
}

@Article{RN1349,
AUTHOR = {Chen, Jiale and Liu, Jianjun and Shu, Fangzhou and Du, Yong and Hong, Zhi},
TITLE = {Merging of Accidental Bound States in the Continuum in Symmetry and Symmetry-Broken Terahertz Photonic Crystal Slabs},
JOURNAL = {Nanomaterials},
VOLUME = {15},
YEAR = {2025},
NUMBER = {6},
ARTICLE-NUMBER = {451},
URL = {https://www.mdpi.com/2079-4991/15/6/451},
PubMedID = {40137624},
ISSN = {2079-4991},
DOI = {10.3390/nano15060451}
}

@article{RN694,
   author = {Gao, Y. and Ge, J. and Gu, Z. and Xu, L. and Shen, X. and Huang, L.},
   title = {Degenerate merging BICs in resonant metasurfaces},
   journal = {Opt Lett},
   volume = {49},
   number = {23},
   pages = {6633-6636},
   ISSN = {1539-4794 (Electronic)
0146-9592 (Linking)},
   DOI = {10.1364/OL.540272},
   url = {https://www.ncbi.nlm.nih.gov/pubmed/39602712},
   year = {2024},
   type = {Journal Article}
}

@article{RN832,
   author = {Heilmann, R. and Salerno, G. and Cuerda, J. and Hakala, T. K. and Torma, P.},
   title = {Quasi-BIC Mode Lasing in a Quadrumer Plasmonic Lattice},
   journal = {ACS Photonics},
   volume = {9},
   number = {1},
   pages = {224-232},
   ISSN = {2330-4022 (Print)
2330-4022 (Electronic)
2330-4022 (Linking)},
   DOI = {10.1021/acsphotonics.1c01416},
   url = {https://www.ncbi.nlm.nih.gov/pubmed/35083367},
   year = {2022},
   type = {Journal Article}
}

@article{RN609,
   author = {Hsu, Chia Wei and Zhen, Bo and Chua, Song-Liang and Johnson, Steven G. and Joannopoulos, John D. and Soljačić, Marin},
   title = {Bloch surface eigenstates within the radiation continuum},
   journal = {Light: Science \& Applications},
   volume = {2},
   number = {7},
   pages = {e84-e84},
   ISSN = {2047-7538},
   DOI = {10.1038/lsa.2013.40},
   year = {2013},
   type = {Journal Article}
}

@article{RN202,
   author = {Hsu, C. W. and Zhen, B. and Lee, J. and Chua, S. L. and Johnson, S. G. and Joannopoulos, J. D. and Soljacic, M.},
   title = {Observation of trapped light within the radiation continuum},
   journal = {Nature},
   volume = {499},
   number = {7457},
   pages = {188-91},
   ISSN = {1476-4687 (Electronic)
0028-0836 (Linking)},
   DOI = {10.1038/nature12289},
   url = {https://www.ncbi.nlm.nih.gov/pubmed/23846657},
   year = {2013},
   type = {Journal Article}
}

@Article{RN194,
author={Hsu, Chia Wei
and Zhen, Bo
and Stone, A. Douglas
and Joannopoulos, John D.
and Solja{\v{c}}i{\'{c}}, Marin},
title={Bound states in the continuum},
journal={Nature Reviews Materials},
year={2016},
month={Jul},
day={19},
volume={1},
number={9},
pages={16048},
issn={2058-8437},
doi={10.1038/natrevmats.2016.48},
url={https://doi.org/10.1038/natrevmats.2016.48}
}

@article{RN150,
   author = {{Hu}, Peng and {Wang}, Jiajun and {Jiang}, Qiao and {Wang}, Jun and {Shi}, Lei and {Han}, Dezhuan and {Zhang}, Z.~Q. and {Chan}, C.~T. and {Zi}, Jian},
        title = "{Global phase diagram of bound states in the continuum}",
      journal = {Optica},
         year = 2022,
        month = dec,
       volume = {9},
       number = {12},
        pages = {1353},
          doi = {10.1364/OPTICA.466190},
       adsurl = {https://ui.adsabs.harvard.edu/abs/2022Optic...9.1353H}
}

@article{RN682,
   author = {Huang, C. and Zhang, C. and Xiao, S. and Wang, Y. and Fan, Y. and Liu, Y. and Zhang, N. and Qu, G. and Ji, H. and Han, J. and Ge, L. and Kivshar, Y. and Song, Q.},
   title = {Ultrafast control of vortex microlasers},
   journal = {Science},
   volume = {367},
   number = {6481},
   pages = {1018-1021},
   ISSN = {1095-9203 (Electronic)
0036-8075 (Linking)},
   DOI = {10.1126/science.aba4597},
   url = {https://www.ncbi.nlm.nih.gov/pubmed/32108108},
   year = {2020},
   type = {Journal Article}
}

@article{RN626,
  title = {Moir\'e Quasibound States in the Continuum},
  author = {Huang, Lei and Zhang, Weixuan and Zhang, Xiangdong},
  journal = {Phys. Rev. Lett.},
  volume = {128},
  issue = {25},
  pages = {253901},
  numpages = {6},
  year = {2022},
  month = {Jun},
  publisher = {American Physical Society},
  doi = {10.1103/PhysRevLett.128.253901},
  url = {https://link.aps.org/doi/10.1103/PhysRevLett.128.253901}
}

@article{RN247,
   author = {Jiang, Q. and Hu, P. and Wang, J. and Han, D. and Zi, J.},
   title = {General Bound States in the Continuum in Momentum Space},
   journal = {Phys Rev Lett},
   volume = {131},
   number = {1},
   pages = {013801},
   ISSN = {1079-7114 (Electronic)
0031-9007 (Linking)},
   DOI = {10.1103/PhysRevLett.131.013801},
   url = {https://www.ncbi.nlm.nih.gov/pubmed/37478422},
   year = {2023},
   type = {Journal Article}
}

@article{RN240,
   author = {Jin, J. and Yin, X. and Ni, L. and Soljacic, M. and Zhen, B. and Peng, C.},
   title = {Topologically enabled ultrahigh-Q guided resonances robust to out-of-plane scattering},
   journal = {Nature},
   volume = {574},
   number = {7779},
   pages = {501-504},
   ISSN = {1476-4687 (Electronic)
0028-0836 (Linking)},
   DOI = {10.1038/s41586-019-1664-7},
   url = {https://www.ncbi.nlm.nih.gov/pubmed/31645728},
   year = {2019},
   type = {Journal Article}
}

@article{RN438,
   author = {Kang, Meng and Liu, Tao and Chan, C. T. and Xiao, Meng},
   title = {Applications of bound states in the continuum in photonics},
   journal = {Nature Reviews Physics},
   volume = {5},
   number = {11},
   pages = {659-678},
   ISSN = {2522-5820},
   DOI = {10.1038/s42254-023-00642-8},
   year = {2023},
   type = {Journal Article}
}

@article{RN698,
   author = {Kang, M. and Mao, L. and Zhang, S. and Xiao, M. and Xu, H. and Chan, C. T.},
   title = {Merging bound states in the continuum by harnessing higher-order topological charges},
   journal = {Light Sci Appl},
   volume = {11},
   number = {1},
   pages = {228},
   ISSN = {2047-7538 (Electronic)
2095-5545 (Print)
2047-7538 (Linking)},
   DOI = {10.1038/s41377-022-00923-4},
   url = {https://www.ncbi.nlm.nih.gov/pubmed/35853861},
   year = {2022},
   type = {Journal Article}
}

@article{RN697,
   author = {Kang, M. and Zhang, S. and Xiao, M. and Xu, H.},
   title = {Merging Bound States in the Continuum at Off-High Symmetry Points},
   journal = {Phys Rev Lett},
   volume = {126},
   number = {11},
   pages = {117402},
   ISSN = {1079-7114 (Electronic)
0031-9007 (Linking)},
   DOI = {10.1103/PhysRevLett.126.117402},
   url = {https://www.ncbi.nlm.nih.gov/pubmed/33798377},
   year = {2021},
   type = {Journal Article}
}

@article{RN146,
   author = {Le, N. D. and Bouteyre, P. and Kheir-Aldine, A. and Dubois, F. and Cueff, S. and Berguiga, L. and Letartre, X. and Viktorovitch, P. and Benyattou, T. and Nguyen, H. S.},
   title = {Super Bound States in the Continuum on a Photonic Flatband: Concept, Experimental Realization, and Optical Trapping Demonstration},
   journal = {Phys Rev Lett},
   volume = {132},
   number = {17},
   pages = {173802},
   ISSN = {1079-7114 (Electronic)
0031-9007 (Linking)},
   DOI = {10.1103/PhysRevLett.132.173802},
   url = {https://www.ncbi.nlm.nih.gov/pubmed/38728718},
   year = {2024},
   type = {Journal Article}
}

@article{RN238,
   author = {Lee, Sun-Goo and Kim, Seong-Han and Kee, Chul-Sik},
   title = {Bound states in the continuum (BIC) accompanied by avoided crossings in leaky-mode photonic lattices},
   journal = {Nanophotonics},
   volume = {9},
   number = {14},
   pages = {4373-4380},
   ISSN = {2192-8614
2192-8606},
   DOI = {10.1515/nanoph-2020-0346},
   year = {2020},
   type = {Journal Article}
}

@article{RN400,
   author = {Li, T. and Wang, J. and Zhang, W. and Wang, X. and Liu, W. and Shi, L. and Zi, J.},
   title = {High-efficiency nonlocal reflection-type vortex beam generation based on bound states in the continuum},
   journal = {Natl Sci Rev},
   volume = {10},
   number = {5},
   pages = {nwac234},
   ISSN = {2053-714X (Electronic)
2095-5138 (Print)
2053-714X (Linking)},
   DOI = {10.1093/nsr/nwac234},
   url = {https://www.ncbi.nlm.nih.gov/pubmed/37113479},
   year = {2023},
   type = {Journal Article}
}

@article{RN248,
   author = {Liu, W. and Wang, B. and Zhang, Y. and Wang, J. and Zhao, M. and Guan, F. and Liu, X. and Shi, L. and Zi, J.},
   title = {Circularly Polarized States Spawning from Bound States in the Continuum},
   journal = {Phys Rev Lett},
   volume = {123},
   number = {11},
   pages = {116104},
   ISSN = {1079-7114 (Electronic)
0031-9007 (Linking)},
   DOI = {10.1103/PhysRevLett.123.116104},
   url = {https://www.ncbi.nlm.nih.gov/pubmed/31573246},
   year = {2019},
   type = {Journal Article}
}

@article{RN467,
author = {Xin Qi and Jiaju Wu and Feng Wu and Mina Ren and Qian Wei and Yufei Wang and Haitao Jiang and Yunhui Li and Zhiwei Guo and Yaping Yang and Wanhua Zheng and Yong Sun and Hong Chen},
journal = {Photon. Res.},
number = {7},
pages = {1262--1274},
publisher = {Optica Publishing Group},
title = {Steerable merging bound states in the continuum on a quasi-flatband of photonic crystal slabs without breaking symmetry},
volume = {11},
month = {Jul},
year = {2023},
url = {https://opg.optica.org/prj/abstract.cfm?URI=prj-11-7-1262},
doi = {10.1364/PRJ.487665},
}

@article{RN1149,
   author = {Rao, L. and Wang, J. and Wang, X. and Wu, S. and Zhao, X. and Liu, W. and Xie, R. and Shen, Y. and Shi, L. and Zi, J.},
   title = {Meron Spin Textures in Momentum Space Spawning from Bound States in the Continuum},
   journal = {Phys Rev Lett},
   volume = {135},
   number = {2},
   pages = {026203},
   ISSN = {1079-7114 (Electronic)
0031-9007 (Linking)},
   DOI = {10.1103/3g3j-mnh9},
   url = {https://www.ncbi.nlm.nih.gov/pubmed/40743166},
   year = {2025},
   type = {Journal Article}
}

@article{RN831,
   author = {Salerno, G. and Heilmann, R. and Arjas, K. and Aronen, K. and Martikainen, J. P. and Torma, P.},
   title = {Loss-Driven Topological Transitions in Lasing},
   journal = {Phys Rev Lett},
   volume = {129},
   number = {17},
   pages = {173901},
   ISSN = {1079-7114 (Electronic)
0031-9007 (Linking)},
   DOI = {10.1103/PhysRevLett.129.173901},
   url = {https://www.ncbi.nlm.nih.gov/pubmed/36332246},
   year = {2022},
   type = {Journal Article}
}

@article{RN296,
author = {Jiajun Wang and Peishen Li and Xingqi Zhao and Zhiyuan Qian and Xinhao Wang and Feifan Wang and Xinyi Zhou and Dezhuan Han and Chao Peng and Lei Shi and Jian Zi},
journal = {Photonics Insights},
number = {11},
pages = {R01},
publisher = {},
title = {Optical bound states in the continuum in periodic structures: mechanisms, effects, and applications},
volume = {3},
month = {1},
year = {2024},
url = {https://www.researching.cn/articles/OJdf2741c1daea1ea5},
}

@article{RN328,
   author = {Wang, J. and Shi, L. and Zi, J.},
   title = {Spin Hall Effect of Light via Momentum-Space Topological Vortices around Bound States in the Continuum},
   journal = {Phys Rev Lett},
   volume = {129},
   number = {23},
   pages = {236101},
   ISSN = {1079-7114 (Electronic)0031-9007 (Linking)},
   DOI = {10.1103/PhysRevLett.129.236101},
   url = {https://www.ncbi.nlm.nih.gov/pubmed/36563232},
   year = {2022},
   type = {Journal Article}
}

@article{RN719,
author = {Wang, Keren and Sun, Kaili and Ding, Qi and Zeng, Lingxiao and Du, Jing and Han, Zhanghua and Huang, Lujun and Wang, Wei},
title = {High-Q Resonance Engineering in Momentum Space for Highly Coherent and Rainbow-Free Thermal Emission},
journal = {Nano Lett.},
volume = {25},
number = {9},
pages = {3613-3619},
year = {2025},
URL = {https://doi.org/10.1021/acs.nanolett.4c06565},
eprint = {https://doi.org/10.1021/acs.nanolett.4c06565}
}

@article{RN292,
author = {Jiaju Wu and Jingguang Chen and Xin Qi and Zhiwei Guo and Jiajun Wang and Feng Wu and Yong Sun and Yunhui Li and Haitao Jiang and Lei Shi and Jian Zi and Hong Chen},
journal = {Photon. Res.},
number = {4},
pages = {638--647},
publisher = {Optica Publishing Group},
title = {Observation of accurately designed bound states in the continuum in momentum space},
volume = {12},
month = {Apr},
year = {2024},
url = {https://opg.optica.org/prj/abstract.cfm?URI=prj-12-4-638},
doi = {10.1364/PRJ.515969},
}

@article{RN246,
   author = {Yoda, T. and Notomi, M.},
   title = {Generation and Annihilation of Topologically Protected Bound States in the Continuum and Circularly Polarized States by Symmetry Breaking},
   journal = {Phys Rev Lett},
   volume = {125},
   number = {5},
   pages = {053902},
   ISSN = {1079-7114 (Electronic)
0031-9007 (Linking)},
   DOI = {10.1103/PhysRevLett.125.053902},
   url = {https://www.ncbi.nlm.nih.gov/pubmed/32794854},
   year = {2020},
   type = {Journal Article}
}

@article{RN736,
   author = {Zhang, N. and Lu, Y. Y.},
   title = {Perturbation Theory for Resonant States near a Bound State in the Continuum},
   journal = {Phys Rev Lett},
   volume = {134},
   number = {1},
   pages = {013803},
   ISSN = {1079-7114 (Electronic)
0031-9007 (Linking)},
   DOI = {10.1103/PhysRevLett.134.013803},
   url = {https://www.ncbi.nlm.nih.gov/pubmed/39913705},
   year = {2025},
   type = {Journal Article}
}

@article{RN436,
   author = {Zhang, Wenjie and Chu, Jiao and Deng, Ruhuan and Wang, Xinhao and Li, Tongyu and Liu, Wenzhe and Wang, Jiajun and Liu, Xiaohan and Shi, Lei},
   title = {Optical Vortices Generation via a Self‐Assembly Photonic Crystal Slab},
   journal = {Advanced Optical Materials},
   volume = {12},
   number = {25},
   ISSN = {2195-1071
2195-1071},
   DOI = {10.1002/adom.202400088},
   year = {2024},
   type = {Journal Article}
}

@article{RN695,
author = {Zhang, Xiao and Xu, JiPeng and Zhang, ZhenBin and Zhu, Zhihong},
title = {Jumping and Merging of Bound States in the Continuum on Degenerate Bands},
journal = {ACS Photonics},
volume = {11},
number = {10},
pages = {4251-4257},
year = {2024},
doi = {10.1021/acsphotonics.4c01164},
URL = {https://doi.org/10.1021/acsphotonics.4c01164
},
eprint = { 
https://doi.org/10.1021/acsphotonics.4c01164
}
}

@article{RN329,
  title = {Spin-Orbit-Locking Chiral Bound States in the Continuum},
  author = {Zhao, Xingqi and Wang, Jiajun and Liu, Wenzhe and Che, Zhiyuan and Wang, Xinhao and Chan, C. T. and Shi, Lei and Zi, Jian},
  journal = {Phys. Rev. Lett.},
  volume = {133},
  issue = {3},
  pages = {036201},
  numpages = {7},
  year = {2024},
  month = {Jul},
  publisher = {American Physical Society},
  doi = {10.1103/PhysRevLett.133.036201},
  url = {https://link.aps.org/doi/10.1103/PhysRevLett.133.036201}
}

@article{RN475,
   author = {Zhen, B. and Hsu, C. W. and Lu, L. and Stone, A. D. and Soljacic, M.},
   title = {Topological nature of optical bound states in the continuum},
   journal = {Phys Rev Lett},
   volume = {113},
   number = {25},
   pages = {257401},
   ISSN = {1079-7114 (Electronic)
0031-9007 (Linking)},
   DOI = {10.1103/PhysRevLett.113.257401},
   url = {https://www.ncbi.nlm.nih.gov/pubmed/25554906},
   year = {2014},
   type = {Journal Article}
}

@article{RN140,
   author = {Zhuang, Z. P. and Zeng, H. L. and Chen, X. D. and He, X. T. and Dong, J. W.},
   title = {Topological Nature of Radiation Asymmetry in Bilayer Metagratings},
   journal = {Phys Rev Lett},
   volume = {132},
   number = {11},
   pages = {113801},
   ISSN = {1079-7114 (Electronic)
0031-9007 (Linking)},
   DOI = {10.1103/PhysRevLett.132.113801},
   url = {https://www.ncbi.nlm.nih.gov/pubmed/38563935},
   year = {2024},
   type = {Journal Article}
}

@inproceedings{Joannopoulos2008PhotonicCM,
  title={Photonic Crystals: Molding the Flow of Light - Second Edition},
  author={John D. Joannopoulos and Steven G. Johnson and Joshua N. Winn and Robert D. Meade},
  year={2008},
  url={https://api.semanticscholar.org/CorpusID:14898450}
}

@Article{RN836,
author={Arjas, Kristian
and Taskinen, Jani Matti
and Heilmann, Rebecca
and Salerno, Grazia
and T{\"o}rm{\"a}, P{\"a}ivi},
title={High topological charge lasing in quasicrystals},
journal={Nature Communications},
year={2024},
month={Nov},
day={05},
volume={15},
number={1},
pages={9544},
issn={2041-1723},
doi={10.1038/s41467-024-53952-5},
url={https://doi.org/10.1038/s41467-024-53952-5}
}

@article{RN1156,
   author = {Chen, Bo and Wei, Yuming and Zhao, Tianming and Liu, Shunfa and Su, Rongbin and Yao, Beimeng and Yu, Ying and Liu, Jin and Wang, Xuehua},
   title = {Bright solid-state sources for single photons with orbital angular momentum},
   journal = {Nature Nanotechnology},
   volume = {16},
   number = {3},
   pages = {302-307},
   ISSN = {1748-3387
1748-3395},
   DOI = {10.1038/s41565-020-00827-7},
   year = {2021},
   type = {Journal Article}
}

@article{RN461,
  title = {Unfolded band structures of photonic quasicrystals and moir\'e superlattices},
  author = {Zhang, Yanbin and Che, Zhiyuan and Liu, Wenzhe and Wang, Jiajun and Zhao, Maoxiong and Guan, Fang and Liu, Xiaohan and Shi, Lei and Zi, Jian},
  journal = {Phys. Rev. B},
  volume = {105},
  issue = {16},
  pages = {165304},
  numpages = {8},
  year = {2022},
  month = {Apr},
  publisher = {American Physical Society},
  doi = {10.1103/PhysRevB.105.165304},
  url = {https://link.aps.org/doi/10.1103/PhysRevB.105.165304}
}

@article{RN322,
   author = {Che, Z. and Zhang, Y. and Liu, W. and Zhao, M. and Wang, J. and Zhang, W. and Guan, F. and Liu, X. and Liu, W. and Shi, L. and Zi, J.},
   title = {Polarization Singularities of Photonic Quasicrystals in Momentum Space},
   journal = {Phys Rev Lett},
   volume = {127},
   number = {4},
   pages = {043901},
   ISSN = {1079-7114 (Electronic)
0031-9007 (Linking)},
   DOI = {10.1103/PhysRevLett.127.043901},
   url = {https://www.ncbi.nlm.nih.gov/pubmed/34355949},
   year = {2021},
   type = {Journal Article}
}

@article{RN1322,
author = {Lei, Xinrui and Zhan, Qiwen},
title = {Topological Quasiparticles of Light: From Spin-Orbit Coupling to Photonic Skyrmions},
journal = {Laser \& Photonics Reviews},
volume = {19},
number = {24},
pages = {e01427},
doi = {https://doi.org/10.1002/lpor.202501427},
url = {https://onlinelibrary.wiley.com/doi/abs/10.1002/lpor.202501427},
eprint = {https://onlinelibrary.wiley.com/doi/pdf/10.1002/lpor.202501427},
year = {2025}
}

@article{RN317,
  title = {Observing vortex polarization singularities at optical band degeneracies},
  author = {Chen, Ang and Liu, Wenzhe and Zhang, Yiwen and Wang, Bo and Liu, Xiaohan and Shi, Lei and Lu, Ling and Zi, Jian},
  journal = {Phys. Rev. B},
  volume = {99},
  issue = {18},
  pages = {180101},
  numpages = {5},
  year = {2019},
  month = {May},
  publisher = {American Physical Society},
  doi = {10.1103/PhysRevB.99.180101},
  url = {https://link.aps.org/doi/10.1103/PhysRevB.99.180101}
}

@article{RN544,
title = {Realization of high-fidelity higher-order Bessel beams},
journal = {Opt. Lasers Eng.},
volume = {184},
pages = {108559},
year = {2025},
issn = {0143-8166},
doi = {https://doi.org/10.1016/j.optlaseng.2024.108559},
url = {https://www.sciencedirect.com/science/article/pii/S0143816624005372},
author = {Chaojie Jiang and Shaohua Tao}
}

@article{RN501,
title = {OAM beam generation in space and its applications: A review},
journal = {Opt. Lasers Eng.},
volume = {151},
pages = {106923},
year = {2022},
issn = {0143-8166},
doi = {https://doi.org/10.1016/j.optlaseng.2021.106923},
url = {https://www.sciencedirect.com/science/article/pii/S0143816621003924},
author = {Yudong Lian and Xuan Qi and Yuhe Wang and Zhenxu Bai and Yulei Wang and Zhiwei Lu}
}

@misc{supp1,
    title = {Supplementary Material: Derivations and Simulation Details},
    note = {This supplementary material covers a $C_4$-symmetric geometric model for off-$\Gamma$ BIC formation, clarifies anisotropic and cross-merging mechanisms (including an intermediate regime), analyzes polarization-resolved anisotropic merging in square-hole PCSs, documents the evolution of SOPs and eigenfields, summarizes reported $\Gamma$-point topological charges, and details numerical methods for high-order momentum-space vortex generation and propagation.},
}

@article{Li2025DegenerateBIC,
  title={Observation of Degenerate Merging Bound States in the Continuum in All-Dielectric Metasurfaces},
  author={Guanhai Li and Chaobiao Zhou and Haoxuan He and others},
  year={2025},
  note={Preprint (Version 1) available at Research Square},
  doi={10.21203/rs.3.rs-7737647/v1},
  url={https://doi.org/10.21203/rs.3.rs-7737647/v1},
  month={October 21},
  publisher={Research Square}
}

@article{Kim2025,
author = {Kim, Dongha and Pelzman, Charles and Guo, Cheng and Long, Olivia and Fan, Shanhui and Cho, Sang},
year = {2025},
month = {09},
pages = {},
title = {Generating topological non-diffracting beams using high quality factor nonlocal metasurfaces},
doi = {10.48550/arXiv.2509.26142},
note={Preprint (Version 1) available at arXiv},
}

@article{wang2025,
      title={Broadband-operational orbital angular momentum generation in nonlocal metasurfaces with maximum efficiency approaching 80\%}, 
      author={Keren Wang and Kaili Sun and Jing Du and Peijuan Dai and Hao Zhou and Lujun Huang and Zhanghua Han and Wei Wang},
      year={2025},
      journal = {arXiv},
      eprint={2510.04131},
      archivePrefix={arXiv},
      primaryClass={physics.optics},
      url={https://arxiv.org/abs/2510.04131}, 
}

@article{RN239,
   author = {Wonjoo, Suh and Zheng, Wang and Shanhui, Fan},
   title = {Temporal coupled-mode theory and the presence of non-orthogonal modes in lossless multimode cavities},
   journal = {IEEE J. Quantum Electron.},
   volume = {40},
   number = {10},
   pages = {1511-1518},
   ISSN = {0018-9197},
   DOI = {10.1109/jqe.2004.834773},
   year = {2004},
   type = {Journal Article}
}

@article{RN631,
doi = {10.1088/1361-6463/acca2e},
url = {https://doi.org/10.1088/1361-6463/acca2e},
year = {2023},
month = {apr},
publisher = {IOP Publishing},
volume = {56},
number = {27},
pages = {275101},
author = {Gao, Wenya and Liu, Ziyi and Li, Xiangning and Wang, Xu and Hu, Guanqu and Ye, Weimin and Guan, Chunying and Liu, Jianlong},
title = {Superimposition of topological charges between vortex beams and singular points of photonic crystal slabs},
journal = {Journal of Physics D: Applied Physics}
}

@article{HUANG20231,
title = {Resonant leaky modes in all-dielectric metasystems: Fundamentals and applications},
journal = {Physics Reports},
volume = {1008},
pages = {1-66},
year = {2023},
note = {Resonant leaky modes in all-dielectric metasystems: Fundamentals and Applications},
issn = {0370-1573},
doi = {https://doi.org/10.1016/j.physrep.2023.01.001},
url = {https://www.sciencedirect.com/science/article/pii/S0370157323000030},
author = {Lujun Huang and Lei Xu and David A. Powell and Willie J. Padilla and Andrey E. Miroshnichenko}
}

@Article{Huang2024,
author={Huang, Lujun
and Huang, Sibo
and Shen, Chen
and Yves, Simon
and Pilipchuk, Artem S.
and Ni, Xiang
and Kim, Seunghwi
and Chiang, Yan Kei
and Powell, David A.
and Zhu, Jie
and Cheng, Ya
and Li, Yong
and Sadreev, Almas F.
and Al{\`u}, Andrea
and Miroshnichenko, Andrey E.},
title={Acoustic resonances in non-Hermitian open systems},
journal={Nature Reviews Physics},
year={2024},
month={Jan},
day={01},
volume={6},
number={1},
pages={11-27},
issn={2522-5820},
doi={10.1038/s42254-023-00659-z},
url={https://doi.org/10.1038/s42254-023-00659-z}
}

@article{advs.202200257,
author = {Huang, Lujun and Jia, Bin and Chiang, Yan Kei and Huang, Sibo and Shen, Chen and Deng, Fu and Yang, Tianzhi and Powell, David A and Li, Yong and Miroshnichenko, Andrey E},
title = {Topological Supercavity Resonances in the Finite System},
journal = {Advanced Science},
volume = {9},
number = {20},
pages = {2200257},
doi = {https://doi.org/10.1002/advs.202200257},
url = {https://advanced.onlinelibrary.wiley.com/doi/abs/10.1002/advs.202200257},
eprint = {https://advanced.onlinelibrary.wiley.com/doi/pdf/10.1002/advs.202200257},
year = {2022}
}

@article{AP.3.1.016004,
author = {Lujun Huang and Lei Xu and Mohsen Rahmani and Dragomir N. Neshev and Andrey E. Miroshnichenko},
title = {{Pushing the limit of high-Q mode of a single dielectric nanocavity}},
volume = {3},
journal = {Advanced Photonics},
number = {1},
publisher = {SPIE},
pages = {016004},
year = {2021},
doi = {10.1117/1.AP.3.1.016004},
URL = {https://doi.org/10.1117/1.AP.3.1.016004}
}

@Article{Huang2021,
author={Huang, Lujun
and Chiang, Yan Kei
and Huang, Sibo
and Shen, Chen
and Deng, Fu
and Cheng, Yi
and Jia, Bin
and Li, Yong
and Powell, David A.
and Miroshnichenko, Andrey E.},
title={Sound trapping in an open resonator},
journal={Nature Communications},
year={2021},
month={Aug},
day={10},
volume={12},
number={1},
pages={4819},
issn={2041-1723},
doi={10.1038/s41467-021-25130-4},
url={https://doi.org/10.1038/s41467-021-25130-4}
}

@article{adfm.202309982,
author = {Huang, Lujun and Li, Shuangli and Zhou, Chaobiao and Zhong, Haozong and You, Shaojun and Li, Lin and Cheng, Ya and Miroshnichenko, Andrey E},
title = {Realizing Ultrahigh-Q Resonances Through Harnessing Symmetry-Protected Bound States in the Continuum},
journal = {Advanced Functional Materials},
volume = {34},
number = {11},
pages = {2309982},
doi = {https://doi.org/10.1002/adfm.202309982},
url = {https://advanced.onlinelibrary.wiley.com/doi/abs/10.1002/adfm.202309982},
eprint = {https://advanced.onlinelibrary.wiley.com/doi/pdf/10.1002/adfm.202309982},
year = {2024}
}

@Article{RN269,
author={Wang, Bo
and Liu, Wenzhe
and Zhao, Maoxiong
and Wang, Jiajun
and Zhang, Yiwen
and Chen, Ang
and Guan, Fang
and Liu, Xiaohan
and Shi, Lei
and Zi, Jian},
title={Generating optical vortex beams by momentum-space polarization vortices centred at bound states in the continuum},
journal={Nature Photonics},
year={2020},
month={Oct},
day={01},
volume={14},
number={10},
pages={623-628},
issn={1749-4893},
doi={10.1038/s41566-020-0658-1},
url={https://doi.org/10.1038/s41566-020-0658-1}
}

@Article{OEA-2023-0186Nicolae,
title = {Resonantly enhanced second- and third-harmonic generation in dielectric nonlinear metasurfaces},
journal = {Opto-Electronic Advances},
volume = {7},
number = {5},
pages = {230186-1-230186-15},
year = {2024},
issn = {2096-4579},
doi = {10.29026/oea.2024.230186},	
url = {https://www.oejournal.org/oea/en/article/doi/10.29026/oea.2024.230186},
author = {Ji Tong Wang and Pavel Tonkaev and Kirill Koshelev and Fangxing Lai and Sergey Kruk and Qinghai Song and Yuri Kivshar and Nicolae C. Panoiu}
}

\end{document}